\documentclass[12pt]{article}
\newcommand{\be}{\begin{equation}}
\newcommand{\ee}{\end{equation}}

\newcommand{\TR}{\mathcal T}
\newcommand{\Par}{\mathcal P}

\newcommand{\bi}{\begin{itemize}}
\newcommand{\ei}{\end{itemize}}
\newcommand{\bea}{\begin{eqnarray}}
\newcommand{\eea}{\end{eqnarray}}
\newcommand{\ba}{\begin{array}}
\newcommand{\ea}{\end{array}}

\usepackage[left=2.50cm, right=2.50cm, top=2.50cm, bottom=2.50cm]{geometry}
\usepackage[utf8]{inputenc}
\usepackage{cancel}
\usepackage[english]{babel}
\usepackage{physics}
\usepackage{hyperref}
\usepackage{amsthm}
\usepackage{mathtools}
\usepackage{graphicx}
\usepackage{changepage}
\usepackage{amssymb,amsmath}
\usepackage{verbatim}
\usepackage{empheq}
\usepackage{xcolor}
\usepackage{tikz}
\usepackage{float}
\usepackage{multirow}
\usepackage{multicol}
\usepackage{amsmath}
\usepackage{centernot}
\usepackage[hang,flushmargin]{footmisc} 
\usepackage[normalem]{ulem}
\usetikzlibrary{tikzmark}
\numberwithin{equation}{section}
\hypersetup{hidelinks}

\newlength{\bibitemsep}
\newlength{\bibparskip}
\let\oldthebibliography\thebibliography
\renewcommand\thebibliography[1]{%
  \oldthebibliography{#1}%
  \setlength{\parskip}{\bibitemsep}%
  \setlength{\itemsep}{\bibparskip}%
}
\begin{document}
\par
\bigskip
\Large
\noindent
{\bf Covariant fracton gauge theory with boundary\\
\medskip
\par
\rm
\normalsize

\hrule

\vspace{.5cm}

\large
\noindent
{\bf Erica Bertolini$^{1,2,a}$}, 
{\bf Nicola Maggiore$^{1,2,b}$},
{\bf Giandomenico Palumbo$^{3,c}$}\\\\

\par
\small
\noindent$^1$ Dipartimento di Fisica, Universit\`a di Genova, via Dodecaneso 33, 16146 Genova, Italy.\\
\smallskip
\noindent$^2$ Istituto Nazionale di Fisica Nucleare (I.N.F.N.) - Sezione di Genova, via Dodecaneso 33, 16146 Genova, Italy.\\
\smallskip
\noindent$^3$ School of Theoretical Physics, Dublin Institute for Advanced Studies, 10 Burlington Road, Dublin 4, Ireland

\smallskip

\vspace{1cm}

\noindent
{\tt Abstract~:}
In this paper we study the consequences of the introduction of a flat boundary on a 4D covariant rank-2 gauge theory described by a linear combination of linearized gravity and covariant fracton theory. We show that this theory gives rise to a Maxwell-Chern-Simons-like theory of two rank-2 traceless symmetric tensor fields. This induced 3D  theory can be physically traced back to the traceless scalar charge theory of fractons, where the Chern-Simons-like term plays the role of a matter contribution.
By further imposing time reversal invariance on the boundary, the Chern-Simons-like term disappears. Importantly, on the boundary of our 4D gauge theory we find a generalized U(1) Ka\c c-Moody algebra and the induced 3D theory is characterized by  the conservation of the dipole moment.

\vspace{\fill}

\noindent{\tt Keywords:} \\
Quantum field theory with boundary, fractons, linearized gravity, symmetric tensor gauge field theory.

\vspace{1cm}

\hrule
\noindent
{\tt E-mail:
$^a$erica.bertolini@ge.infn.it,
$^b$nicola.maggiore@ge.infn.it,\\
$^c$giandomenico.palumbo@gmail.com.
}
\newpage

\section{Introduction }

Fracton phases of matter represent a novel paradigm both in condensed matter theory and high energy physics \cite{Nandkishore:2018sel,Pretko:2020cko}. Although they have originally been discovered in particular kinds of lattice models \cite{Chamon:2004lew,Haah:2011drr,Vijay:2016phm,Vijay:2015mka}, fractons have been unveiled in many different systems and frameworks, ranging from elasticity \cite{Pretko:2017kvd,Gromov,Musso,Moroz}, hydrodynamics \cite{Lucas,Ye,Wang,Surowska2,Surowska3} and quantum scars \cite{Pai,Khemani,Sala}, to Quantum Field Theories (QFTs) \cite{Seiberg:2020wsg,Pretko:2016lgv,Pretko:2016kxt,Bulmash:2018lid,Shirley:2017suz,Ma:2017aog,Ma:2018nhd,Williamson,Slagle,Slagle2,Slagle3}, curved space \cite{Prem,Bidussi,Jain,Tsaloukidis:2023bvz}
and holography \cite{Yan}. These phases are characterized by constrained dynamics, for which quasiparticle excitations are immobile or move in sub-dimensional spaces. This exotic behaviour can be encoded into conservation of multipole moment, the simplest example being the dipole. Indeed,  typically fracton models are described in terms of a non-covariant higher-rank tensor theory which shares many similarities with Maxwell theory \cite{Pretko:2016lgv,Pretko:2016kxt,Pretko:2017xar}. They are written in terms of a rank-2 symmetric tensor field $A_{ij}(x)$, whose conjugate momentum is referred to as an ``electric-like'' tensor field $E^{ij}(x)$, which plays a key role for the immobility constraint of fractons, which is recovered from a generalized Gauss law. For instance in the so called ``scalar charge theory of fractons'' \cite{Pretko:2016lgv,Pretko:2016kxt} this law appears as $ \partial_i\partial_jE^{ij}=\rho$, which implies dipole ($x^i\rho$) conservation, which, in turn, implies that single charges cannot move in isolation. Other possibilities are also allowed~: for instance one can consider a ``vector charge'' model, where the nature of the charge changes, becoming a vector $\rho^i(x)$, while the ``electric field'' is always a symmetric rank-2 tensor. In that case the constraint involves one derivative instead of two $\partial_jE^{ij}=\rho^i$ and the conservation concerns linear and angular momenta, which implies that the charges of the system can move only along a line, thus being 1-dimensional particles (also called ``lineons''). Mobility can be further restricted in both scalar and vector charge models by adding tracelessness $E^i_{\,i}=0$ as an additional constraint. In these cases the models are referred to as ``traceless (scalar/vector) fracton models''. In the scalar case the elementary charges still can be identified as fractons, since their motion is already maximally constrained by means of the generalized Gauss law, however now tracelessness also implies the conservation of a component of the quadrupole momentum, due to which the dipoles of the system are bound to move only on a plane transverse to their direction, thus becoming 2-dimensional particles (also called ``planons''). On the contrary, in the vector charge model the quasiparticles can move in one dimension, and the tracelessness constraint on the electric-like field completely restricts their motion
making them proper fractons. These novel gauge theories are intrinsically non-relativistic, and many ingredients have been introduced by hand in order to implement the main characteristics of fractons, $i.e. $ their restricted mobility. Examples of these inputs are, for instance, the Maxwell-like Hamiltonian and the Gauss law, introduced as an external constraint, and not derived from an action principle. As a consequence, terms appear with inhomogeneous numbers of derivatives, as remarked in \cite{Brauner:2022rvf}, and all these $ad\ hoc$ introductions are justified $a\ posteriori$, rather than deduced from first principles of QFT. Despite their intrinsic non-covariance, these models share remarkable similarities both with Maxwell theory (Gauss law, Hamiltonian, electric-like field...) \cite{Pretko:2016lgv} and Linearized Gravity (LG) (symmetric rank-2 field, gauge symmetry,...) \cite{Pretko:2017fbf}, which however are fully covariant theories. Motivated by these similarities, in \cite{Blasi:2022mbl,Bertolini:2022ijb,Bertolini:2023juh} a covariant 4D fracton gauge theory has been built, taking as unique ingredients locality, power counting and the covariant fracton symmetry $\delta A_{\mu\nu}=\partial_\mu \partial_\nu \Lambda$. This made immediately apparent the correspondence with LG, and all the analogies with Maxwell theory naturally came out.
Embedding the ordinary non-covariant theory of fractons in a larger covariant theory led to recover all the known results concerning fractons \cite{Pretko:2016lgv,Pretko:2016kxt,Pretko:2017xar}. Therefore, the coherent theoretical framework of \cite{Bertolini:2022ijb} allows to apply standard QFT techniques to fractons. As a non-trivial and physically relevant example, in this paper we asked the question of which might be the consequences of the introduction of a flat spatial boundary in the 4D covariant fracton model \cite{Bertolini:2022ijb}. In fact, the  introduction of boundaries in QFT is  known to give rise to rich physical results. The most notable examples come from Topological Field Theories \cite{Birmingham:1991ty}, which represent the effective field theories of the bulk states of topological phases of matter. In fact, when a boundary is present, Chern-Simons (CS) theories in 3D, $\theta$-terms in 4D and BF models in any spacetime dimensions give rise to the theoretical descriptions of the edge states of quantum Hall fluids and topological insulators, respectively \cite{Zhang:1988wy,Qi,Cho:2010rk,Cirio}. However also in non-topological field theories the presence of a boundary has non-trivial consequences. For instance on the boundary of Maxwell theory in 3D and 4D \cite{Maggiore:2019wie,Bertolini:2020hgr} a Ka\c{c}-Moody (KM) algebra is observed, and a non-trivial theory is induced on the lower-dimensional space. Similarly, the topological $\theta$-term \cite{tong}, which is a boundary term, 
generates the well known Witten effect \cite{Witten:1979ey}, relevant for instance in topological insulators \cite{Rosenberg:2010ia}. 
For what concerns fractons, boundary contributions have been introduced mainly as non-covariant CS-like terms coming from a generalized topological-like $\theta$-term in the bulk \cite{Pretko:2017xar,Burnell,Burnell2}, inspired by the standard electromagnetic case. For instance in \cite{Pretko:2017xar}, as for the Maxwell case, a Witten-like effect is observed, for which, due to the presence of the fractonic $\theta$-term, the ``electric''  charge density at the right hand side of the Gauss law acquires an additional ``magnetic''  contribution. Furthermore on the 3D boundary of \cite{Pretko:2017xar} fractonic-like excitations seem to appear, in agreement with our results, as we shall see. 
Moreover, in \cite{Pretko:2020cko,You:2019bvu} it has been speculated that certain kinds of higher-order topological phases share some properties with fracton quasiparticles.
Another interesting example can be found in \cite{Cappelli:2015ocj}, where a similar non-covariant CS-like term, built as a higher-spin generalization of the standard topological one, gives some insights in the context of dipolar behaviours of quantum Hall systems.
Therefore, boundary effects might be important also in the framework of fracton theory, and the aim of this paper is to study the consequences of the presence of a flat boundary in the covariant theory of fractons, following the QFT approach pioneered by Symanzik in \cite{Symanzik:1981wd}.\\

The paper is organized as follows: in Section \ref{sec-model} we briefly review the 4D covariant fracton theory of \cite{Bertolini:2022ijb}, where the most general action invariant under the covariant fracton symmetry $\delta A_{\mu\nu}=\partial_\mu\partial_\nu\Lambda$ is identified as the combination of two independent terms: LG and pure fractons. In Section \ref{sec-model+bd} the boundary is introduced in the action together with the gauge fixing and the most general boundary term. From the total action the equations of motion (EoM) and the most general boundary conditions (BC) are computed. Because of the presence of the boundary, the Ward identities of the theory are broken, and this allows to identify the boundary degrees of freedom (DoF), 
represented by two traceless symmetric rank-2 tensor fields. Moreover, the broken Ward identities give rise to an algebraic structure, which can be identified as a generalized KM algebra, which, in Section \ref{sec-3Dmodel}, we interpret as canonical commutators of a 3D action. In Section \ref{sec-hc} the bulk/boundary correspondence is obtained by requiring that the EoM of the induced 3D action are compatible with the boundary conditions of the 4D bulk theory. This can be achieved by suitably tuning the parameters appearing in the boundary conditions and in the 3D action. To physically interpret the 3D theory we found, 
in Section \ref{sec-phys} we study its EoM, which appear to be 
Gauss and Amp\`ere-like laws for the boundary tensor fields, exactly as in the ``ordinary'' fracton theory. In particular our boundary theory can be identified with a
traceless fracton model. In Section \ref{sec PT} we analyze the effect of taking into account discrete symmetries such as parity ($\Par$)  and time reversal ($\TR$). Finally in Section \ref{sec-summary} we discuss our results.

\normalcolor

\section{The model without boundary: fractons and linearized gravity}\label{sec-model}

Following \cite{Blasi:2022mbl,Bertolini:2022ijb,Bertolini:2023juh}, we start by considering the covariant extension of the fracton symmetry in the scalar charge theory
\begin{equation}\label{dA}
	\delta_{fract} A_{\mu\nu}=\partial_\mu\partial_\nu\Lambda\ ,
\end{equation}
which identifies the most general 4D invariant action
\be
S_{inv}=g_1S_{fract}+g_2S_{LG}\ ,
\label{Sinvg1g2}\ee
where
\bea
S_{fract} &=&
\frac{1}{6}\;\int d^4x\;F^{\mu\nu\rho}F_{\mu\nu\rho}
\label{SfractF}\\
S_{LG} &=& 
\int d^4x\; \left(
\frac{1}{4}F^\mu_{\ \mu\nu} F_\rho^{\ \rho\nu}-\frac{1}{6}F^{\mu\nu\rho}F_{\mu\nu\rho}
\right)\ ,
\label{SLGF}
\eea
 $g_1,g_2$ are dimensionless constants and $F_{\mu\nu\rho}(x)$ is the covariant fracton field strength, defined as \cite{Bertolini:2022ijb,Wu:1988py}
	\begin{equation}
	F_{\mu\nu\rho}=F_{\nu\mu\rho}=\partial_\mu A_{\nu\rho}+\partial_\nu A_{\mu\rho}-2\partial_\rho A_{\mu\nu}\ ,
	\label{Fmunurho}\end{equation}
which has the following properties, compared with those of the ordinary Maxwell field strength
\begin{table}[H]
\centering
  \begin{tabular}{ | l | c | c| }
    \hline
     & fractons & Maxwell \\ \hline
      invariance & $\delta_{fract}F_{\mu\nu\rho}=0$& $\delta_{gauge}F_{\mu\nu}=0$ \\ \hline
    cyclicity & $F_{\mu\nu\rho}+F_{\nu\rho\mu}+F_{\rho\mu\nu}=0$ & $F_{\mu\nu}+F_{\nu\mu}=0$ \\ \hline
    Bianchi & $\epsilon_{\alpha\mu\nu\rho}\partial^{\mu}F^{\beta\nu\rho}=0$ & 
   $ \epsilon_{\mu\nu\rho\sigma}\partial^\nu F^{\rho\sigma}=0$ \\
    \hline
  \end{tabular}
  \caption{\footnotesize\label{table1} Properties of the fracton and Maxwell field strengths.}
\end{table}
The covariant symmetry \eqref{dA} makes apparent the relation between fracton theories and gravity, as already guessed in \cite{Pretko:2017fbf,Casini,Pena-Benitez}. In fact $S_{LG}$ in \eqref{Sinvg1g2}, obtained at $g_1=0$, is the action of LG written in terms of $F_{\mu\nu\rho}(x)$ \cite{Bertolini:2022ijb}, which is defined by the more general infinitesimal diffeomorphism symmetry
\be\label{diff}
\delta_{diff}A_{\mu\nu}=\partial_\mu\xi_\nu+\partial_\nu\xi_\mu\ ,
\ee
of which  \eqref{dA} is a particular case, sometimes called longitudinal diffeomorphisms \cite{Dalmazi:2020xou}.  The fracton action $S_{fract}$ \eqref{SfractF}, instead,  corresponds to the limit $g_2=0$, which has been shown in \cite{Bertolini:2022ijb} to yield all the results associated to the scalar charge theory of fractons \cite{Pretko:2016kxt,Pretko:2016lgv,Pretko:2017xar,Nandkishore:2018sel,Pretko:2020cko}. From the action $S_{inv}$ \eqref{Sinvg1g2} we get the EoM
\be
\frac {\delta S_{inv}}{\delta A^{\alpha\beta}} =g_1\partial^{\mu}F_{\alpha\beta\mu}
+g_2\left[\eta_{\alpha\beta}\partial_\mu F_\nu^{\ \nu\mu}-\frac{1}{2}\left(\partial_\alpha F^\mu_{\ \mu\beta}+\partial_\beta F^{\mu}_{\ \mu\alpha}\right)-\partial^{\mu}F_{\alpha\beta\mu}\right]=0\ ,
\label{eom inv}
\ee
and  the conjugate momentum
\be
\Pi^{\alpha\beta}(g_1,g_2)\equiv\frac{\partial\mathcal L_{inv}}{\partial(\partial_tA_{\alpha\beta})}
=
-g_1 F^{\alpha\beta0}
-g_2\left[
\eta^{\alpha\beta}F_{\lambda}^{\ \lambda0}-\frac{1}{2}\left(\eta^{0\alpha}F_{\lambda}^{\ \lambda\beta}+\eta^{0\beta}F_{\lambda}^{\ \lambda\alpha}\right) -F^{\alpha\beta0}\right]\ ,
\label{E^ij inv}\ee
which in fracton theories plays a relevant role, since its spatial components are identified with the ``tensor electric field'' $E^{ij}(x)$ \cite{Nandkishore:2018sel,Pretko:2020cko,Pretko:2016lgv,Pretko:2016kxt,Pretko:2017xar}. The components of $\Pi^{\alpha\beta}(x)$ are
\bea
	\Pi^{00}&=&0 \label{Pi00=0}\\
	\Pi^{i0}&=&-g_1F^{i00}-\frac{1}{2}g_2 F_j^{\ ji}\label{Pi0i}\\
	\Pi^{ij}&=&-g_1F^{ij0} +g_2(F^{ij0}-\eta^{ij}F_k^{\ k0}) \label{Piij}\ ,
\eea
with $i,j=\{1,2,3\}$. Notice that for a particular combination of $g_1$ and $g_2$ the trace of $\Pi^{\alpha\beta}$ vanishes
\be
\eta_{\alpha\beta}\Pi^{\alpha\beta}
=\Pi^\alpha_{\ \alpha}=-(g_1+2g_2)F_{\lambda}^{\ \lambda0}=0\quad \mbox{if $g_1+2g_2=0$}\ .
\label{tracelessconstr}\ee
This corresponds to the fact that, as already remarked in \cite{Blasi:2022mbl}, in this case the action $S_{inv}$ \eqref{Sinvg1g2} does not depend on the trace of the tensor field $A_{\mu\nu}(x)$, thus further lowering the  number of DoF \cite{Bertolini:2023juh}.

\section{The model with boundary}\label{sec-model+bd}

\subsection{The action}
In view of introducing a planar boundary $x^3=0$, we use the following conventions concerning indices :
	\begin{empheq}{align}
	\alpha,\beta,\gamma,...&=\{0,1,2,3\}\\
	a,b,c,...&=\{0,1,2\}\\
	\textsc{a,b,c},...&=\{1,2\}\ .
	\end{empheq}
	Moreover, $x^\mu=(x^0,x^1,x^2,x^3)$ and $X^m=(x^0,x^1,x^2)$ are the bulk and boundary coordinates respectively. We now introduce the boundary by means of a Heaviside step function in the action \cite{Amoretti:2014iza}:
	\be\label{Sbulk}
	S_{bulk} =
	\int d^4x\;\theta(x^3)\left\{\frac{g_1}{6}F^{\mu\nu\rho}F_{\mu\nu\rho}+g_2 \left(
	\frac{1}{4}F^\mu_{\ \mu\nu} F_\rho^{\ \rho\nu}-\frac{1}{6}F^{\mu\nu\rho}F_{\mu\nu\rho}
	\right)\right\}\ .
	\ee
Notice that in what follows we cannot just set $g_1=0$ and restrict our results to LG alone, because 
 $S_{LG}$ \eqref{SLGF} is uniquely defined by the infinitesimal diffeomorphism transformation \eqref{diff}, and not by its subset \eqref{dA}.
The transformations \eqref{diff} and \eqref{dA} differs in two aspects: the first, \eqref{diff}, depends on a vector gauge parameter, while \eqref{dA} has a scalar gauge parameter, hence the former is more restrictive. Secondly, \eqref{diff} and \eqref{dA} depend on one and two derivatives respectively. This results in a mismatch in the mass dimensions. In fact, since from the action \eqref{Sbulk} we have $[A_{\mu\nu}]=1$, due to the double derivatives in \eqref{dA} it must be $[\Lambda]=-1$, which is an exotic dimension assignment for the scalar gauge parameter. Moreover, on the $x^3$-boundary, the field $A_{\mu\nu}(x)$ and its $\partial_3$-derivative must be treated as independent fields \cite{Maggiore:2019wie,Bertolini:2020hgr,Karabali:2015epa,Blasi:2019wpq}. Hence, on the boundary we define
	\be
	\tilde A_{\mu\nu}\equiv\partial_3A_{\mu\nu}|_{x^3=0}\ ,
	\ee
with $[\tilde A_{\mu\nu}]=2$.
We add to the invariant action $S_{inv}$ \eqref{Sinvg1g2} the gauge-fixing term 
	\be
	S_{gf}=\int d^4x\;\theta(x^3)b^\mu A_{\mu3}\ ,
	\ee
where $b^\mu(x)$ is a Nakanishi-Lautrup Lagrange multiplier \cite{Nakanishi:1966zz,Lautrup:1967zz} implementing the axial gauge condition 
	\be\label{ax-gf}
	A_{\mu3}=0\ .
	\ee
As a consequence of the fact that the field and its $\partial_3$-derivative on the boundary are independent quantities, together with the usual external field $J^{ab}(x)$ coupled to $A_{ab}(x)$, it is necessary to couple a source $\tilde J^{ab}(x)$ also to the $\partial_3$-derivative of $A_{ab}(x)$ on the boundary. The external source term is then
	\be
	S_J=\int d^4x\left[\theta(x^3)J^{ab}A_{ab}+\delta(x^3)\tilde J^{ab}\tilde A_{ab}\right]\ .
	\ee
	The presence of a boundary in a QFT naturally rises the question of which BC should be assigned to the quantum fields and/or their derivatives. A possible way is to impose them by hand, but one should worry about the dependence of the results on the particular choice. This arbitrariness affecting QFTs with boundary has been elegantly solved by Symanzik in his pioneering paper \cite{Symanzik:1981wd}, where a scalar QFT with boundary was considered. According to Symanzik's approach, the BC are not imposed by hand, but are determined by the theory itself. This is achieved by adding a boundary term to the action, as the most general one, satisfying the requests of locality, power counting and 3D Lorentz invariance. The BC are then determined from the EoM, modified by the boundary term
	\be\label{Sbd}
	S_{bd}=\int d^4x\delta(x^3)\left[\xi_0 A_{ab}A^{ab}+\xi_1\tilde A_{ab}A^{ab}+\xi_2\epsilon^{abc}A_{ai}\partial_bA_c^i+\xi_3A^2+\xi_4\tilde AA\right]\ ,
	\ee
where, due to the gauge condition \eqref{ax-gf},
	\be
	A\equiv\eta^{\mu\nu}A_{\mu\nu}=\eta^{ab}A_{ab}\quad;\quad\tilde A\equiv\eta^{\mu\nu}\tilde A_{\mu\nu}=\eta^{ab}\tilde A_{ab}\ ,
	\ee
and $\xi_i$ are constant parameters, whose mass dimensions are
	\be
	[\xi_0]=[\xi_3]=1\quad;\quad[\xi_1]=[\xi_2]=[\xi_4]=0\ .
	\ee
Notice that the general QFT requirements which constrain $S_{bd}$ imply the presence of the CS-like $\xi_2$-term, which can be traced back to the covariant fractonic $\theta$-term \cite{Bertolini:2022ijb}. In fact, this latter is 
\be
S_\theta=\int d^4x \theta(x^3) \epsilon^{\mu\nu\rho\sigma} \partial_\mu A_{\nu\lambda}\partial_\rho A_\sigma^\lambda\ ,
\label{thetaterm}\ee
which, integrating by parts, reduces to
\be
\int d^3X \epsilon^{abc} A_{a\lambda}\partial_b A_c^\lambda\ ,
\label{xi2term}\ee
which, on the gauge condition \eqref{ax-gf}, coincides with the $\xi_2$-term in \eqref{Sbd}.
The total action is then
	\be\label{Stot}
	S_{tot}=S_{bulk}+S_{gf}+S_{J}+S_{bd}\ .
	\ee
	
\subsection{Equations of motion and boundary conditions}

The EoM for $A_{\alpha\beta}(x)$ and $\tilde A_{\alpha\beta}(x)$ are
\be
\begin{split}
\frac {\delta S_{tot}}{\delta A_{\alpha\beta}} =&\theta(x^3)\left\{(g_1-g_2)\partial_{\mu}F^{\alpha\beta\mu}
+g_2\left[\eta^{\alpha\beta}\partial_\mu F_\nu^{\ \nu\mu}-\tfrac{1}{2}\left(\partial^\alpha F_\mu^{\ \mu\beta}+\partial^\beta F_{\mu}^{\ \mu\alpha}\right)\right]+\delta^\alpha_a\delta^\beta_bJ^{ab}+\tfrac{1}{2}(b^\alpha\delta^\beta_3+b^\beta\delta^\alpha_3)\right\}+\\
&+\delta(x^3)\left\{(g_1-g_2)F^{\alpha\beta3}+g_2\left[\eta^{\alpha\beta}F_\mu^{\ \mu3}-\tfrac{1}{2}\left(\eta^{\alpha3}F_\mu^{\ \mu\beta}+\eta^{\beta3}F_\mu^{\ \mu\alpha}\right)\right]+\right.\\
&+\left.\delta^\alpha_a\delta^\beta_b\left[ 2\xi_0A^{ab}+\xi_1\tilde A^{ab}+\xi_2(\epsilon^{aij}\partial_iA_j^b+\epsilon^{bij}\partial_iA_j^a)+2\xi_3\eta^{ab}A+\xi_4\eta^{ab}\tilde A\right]\right\}=0\ ,\label{eomA}
\end{split}
\ee
and
\be
\begin{split}
\frac {\delta S_{tot}}{\delta \partial_3A_{\alpha\beta}} =&\theta(x^3)\left\{(g_2-g_1)F^{\alpha\beta3}
-g_2\left[\eta^{\alpha\beta}F_\mu^{\ \mu3}-\tfrac{1}{2}\left(\eta^{\alpha3}F_\mu^{\ \mu\beta}+\eta^{\beta3}F_{\mu}^{\ \mu\alpha}\right)\right]\right\}\\
&+\delta(x^3)\delta^\alpha_a\delta^\beta_b\left\{ \tilde J^{ab}+\xi_1A^{ab}+\xi_4\eta^{ab}A\right\}=0\ .\label{eomAt}
\end{split}
\ee	
The most general BC are obtained by applying  $\lim_{\epsilon\to0}\int^\epsilon_0dx^3$ to the EoM. From \eqref{eomA} we get
\be
\begin{split}
&\left\{(g_1-g_2)F^{\alpha\beta3}+g_2\left[\eta^{\alpha\beta}F_\mu^{\ \mu3}-\tfrac{1}{2}\left(\eta^{\alpha3}F_\mu^{\ \mu\beta}+\eta^{\beta3}F_\mu^{\ \mu\alpha}\right)\right]+\right.\\
&+\left.\delta^\alpha_a\delta^\beta_b\left[ 2\xi_0A^{ab}+\xi_1\tilde A^{ab}+\xi_2(\epsilon^{aij}\partial_iA_j^b+\epsilon^{bij}\partial_iA_j^a)+2\xi_3\eta^{ab}A+\xi_4\eta^{ab}\tilde A\right]\right\}_{x^3=0}=0\ .\label{bcA}
\end{split}
\ee
We observe that
\begin{itemize}
\item $\alpha=\beta=3$ is trivially realized ;
\item $\alpha=3,\ \beta=b$ :
	\be
	g_2\left(\partial^bA-\partial_aA^{ab}\right)_{x^3=0}=g_2F_\mu^{\ \mu b}|_{x^3=0}=0\ ;\label{bcA3i}
	\ee
\item $\alpha=a,\ \beta=b$ :
	\be
	\left[ 2\xi_0A^{ab}+2(g_2-g_1+\frac{\xi_1}{2})\tilde A^{ab}+\xi_2(\epsilon^{aij}\partial_iA_j^b+\epsilon^{bij}\partial_iA_j^a)+2\xi_3\eta^{ab}A+2(\frac{\xi_4}{2}-g_2)\eta^{ab}\tilde A\right]_{x^3=0}=0\ .\label{bcAij}
	\ee
\end{itemize}
Going on-shell, $i.e.$ at vanishing external sources $\tilde J(x)=0$, taking the $\lim_{\epsilon\to0}\int^\epsilon_0dx^3$ of the EoM \eqref{eomAt} we get
\be
\delta^\alpha_a\delta^\beta_b\left(\xi_1A^{ab}+\xi_4\eta^{ab}A\right)_{x^3=0}=0\ ,\label{bcAt}
\ee
and again we observe that
\begin{itemize}
\item $\alpha=3,$ $\beta$ free, is trivially realized ;
\item $\alpha=a,\ \beta=b$ :
\be
\left(\xi_1A^{ab}+\xi_4\eta^{ab}A\right)_{x^3=0}=0\ .\label{bcAtij}
\ee
\end{itemize}

\subsection{Ward identities and boundary degrees of freedom}

The EoM \eqref{eomA} yields the following integrated Ward identity
	\be
	\int dx^3\partial_a\partial_b\frac{\delta S_{tot}}{\delta A_{ab}}
	=\int dx^3\theta(x^3)\left\{(g_1-g_2)\partial_a\partial_b\partial_{3}F^{ab3}+g_2\partial_a\partial^a \partial_3F_\mu^{\ \mu 3}+\partial_a\partial_bJ^{ab}\right\}=0\ ,
	\ee
where we used the BC \eqref{bcAij} and the cyclic property of $F_{\mu\nu\rho}(x)$ in Table \ref{table1}. Integrating by parts we get
	\be\label{wi1}
	\int dx^3\theta(x^3)\partial_i\partial_jJ^{ij}=
	2(g_2-g_1)\partial_i\partial_j\tilde A^{ij}-2g_2\partial_i\partial^i\tilde A|_{x^3=0}\ .
	\ee
Analogously, from the EoM \eqref{eomAt} we find
	\be
	\int dx^3\partial_a\partial_b\frac{\delta S_{tot}}{\delta \partial_3A_{ab}}
	=-\int dx^3\theta(x^3)\partial_a\partial_b\left[2(g_2-g_1)\partial^3A^{ab}-2g_2\eta^{ab}\partial^3A\right]+\partial_a\partial_b\tilde J^{ab}|_{x^3=0}=0\ ,
	\ee
where we used the BC  \eqref{bcAtij}. Integrating by parts
	\be\label{wi2}
	\partial_i\partial_j\tilde J^{ij}|_{x^3=0}=-2(g_2-g_1)\partial_i\partial_j A^{ij}+2g_2\partial_i\partial^iA|_{x^3=0}\ .
	\ee
	Notice that the second Ward identity \eqref{wi2}, associated to the $\tilde A_{ab}(x)$ field on the boundary, is local and not integrated as the first \eqref{wi1}. The two Ward identities \eqref{wi1} and \eqref{wi2} are analogous to those characterizing Maxwell theory with boundary both in 3D and 4D \cite{Maggiore:2019wie,Bertolini:2020hgr}. At vanishing external source $J^{ab}(x)=0$, the Ward identity \eqref{wi1} gives
	\be
	\partial_i\partial_j\left[(g_2-g_1)\tilde A^{ij}-g_2\eta^{ij}\tilde A\right]_{x^3=0}=0\ .\label{cc1}
	\ee
In \cite{Blasi:2022mbl, Bertolini:2023juh} it has been shown that, when $g_1=g_2$ in the invariant action $S_{inv} $ \eqref{Sinvg1g2}, it is possible to redefine the components of $A_{\mu\nu}(x)$ in such a way that the theory has no kinetic term, hence it is not dynamical. Therefore in what follows we shall exclude the trivial case
\be
g_1=g_2\ .
\ee
For $g_1\neq g_2$, and $g_{1,2}\neq0$, \eqref{cc1} implies
	\begin{empheq}{align}
	\partial_i\partial_j\tilde A^{ij}(X)&=0\label{ddA=0}\\
	\Box \tilde A(X)&=0\ .\label{boxA=0}
	\end{empheq}
Eq.\eqref{ddA=0} is solved as follows \cite{Henneaux:2004jw,Bunster:2012km}
	\be
	\partial_i\left(\partial_j\tilde A^{ij}\right)=0\quad\Rightarrow\quad \partial_j\tilde A^{ij}=\epsilon^{imn}\partial_mC_n\ ,\label{dA=Cn}
	\ee
where $C_n(X)$ is a generic 3D vector field. Eq.\eqref{dA=Cn}, in turn, gives
	\be
	\partial_j\left(\tilde A^{ij}-\epsilon^{ijn}C_n\right)=0\quad\Rightarrow\quad \tilde A^{ij}-\epsilon^{ijn}C_n=2\epsilon^{jab}\partial_a\tilde a^{\ i}_{b}\ ,
	\ee
where $\tilde a_{ij}(X)$ is a generic rank-2 tensor field. On the other hand, $\tilde A^{ij}(x)$ is symmetric, hence $C_n=0$ and  we have
\normalcolor
	\be\label{sol1}
	\tilde A^{ij}(X)\equiv\epsilon^{iab}\partial_a\tilde a_b^{\ j}(X)+\epsilon^{jab}\partial_a\tilde a_b^{\ i}(X)\ .
	\ee
The tensor field $\tilde a_{ij}(X)$ represents the DoF on the boundary, with $[\tilde a_{ij}]=1$. Moreover, since $\tilde A_{ij}(x)=\tilde A_{ji}(x)$ has six independent components, the boundary field $\tilde a_{ij}(X)$ must be symmetric as well
	\be
	\tilde a_{ij}=\tilde a_{ji}\ ,
	\ee
in order that the boundary DoF does not exceed the number of components of its bulk ancestor $\tilde A_{ij}(x)$. The solution \eqref{sol1} is traceless 
	\be
	\tilde A(X)|_\eqref{sol1}=0\ ,
	\ee
so that the condition  \eqref{boxA=0} is automatically satisfied. 
Analogously, from the local Ward identity \eqref{wi2} we have, at vanishing external source $\tilde J^{ab}(x)=0$
	\be
	\partial_i\partial_j\left[(g_2-g_1) A^{ij}-g_2\eta^{ij} A\right]_{x^3=0}=0\ ,\label{cc2}
	\ee
whose solution is
	\be\label{sol2}
	 A^{ij}(X)\equiv\epsilon^{iab}\partial_a a_b^{\ j}(X)+\epsilon^{jab}\partial_a a_b^{\ i}(X)\ ,
	\ee
where $a_{ij}(X)=a_{ji}(X)$ is the DoF on the boundary with $[a_{ij}]=0$. Let us now consider the two broken Ward identities \eqref{wi1} and \eqref{wi2} and make functional derivatives with respect to $J^{mn}(X')$ and $\tilde J^{mn}(X')$. Referring to Appendix \ref{appAlgebra} for the details, we obtain the following equal time commutation relations
	\begin{empheq}{align}
	\left[\Delta\tilde A(X)\ ,\ A_{\textsc{mn}}(X')\right]_{x^0=x'^0}&=+i\partial_\textsc{m}\partial_\textsc{n}\{\delta(x^1-x'^1)\delta(x^2-x'^2)\}\label{[DAt,A]}\\
	\left[\Delta A(X)\ ,\ \tilde A_{\textsc{mn}}(X')\right]_{x^0=x'^0}&=-i\partial_\textsc{m}\partial_\textsc{n}\{\delta(x^1-x'^1)\delta(x^2-x'^2)\}\label{[DA,At]}\ ,
	\end{empheq}
where
	\begin{empheq}{align}
	\Delta \tilde A&\equiv2(g_1-g_2)\left(\partial_j\tilde A^{0j}+\partial_\textsc{a}\tilde A^{0\textsc{a}}\right)+2g_2\partial^0\tilde A\label{DAt}\\	
	\Delta A&\equiv2(g_1-g_2)\left(\partial_jA^{0j}+\partial_\textsc{a}A^{0\textsc{a}}\right)+2g_2\partial^0A\ .\label{DA}
	\end{empheq}
Importantly, the commutation relations \eqref{[DAt,A]} and \eqref{[DA,At]} resemble the generalized U(1) KM algebra derived in \cite{You:2019bvu} for a 3D non-chiral bosonic theory that lives on the boundary of a 4D dipolar fracton theory. Hence, the theory described by the action $S_{inv}$ \eqref{Sinvg1g2} has an algebraic structure on the boundary that is different from that of topological field theories \cite{Blasi:2019wpq,Blasi:2008gt,Blasi:2010gw,Blasi:2011pf,Amoretti:2012hs,Amoretti:2013nv,Amoretti:2013xya,Amoretti:2014kba,Amoretti:2014iza,Maggiore:2017vjf,Maggiore:2018bxr,Bertolini:2021iku,Bertolini:2022sao} and Maxwell theory \cite{Maggiore:2019wie,Bertolini:2020hgr,Iqbal}. The reason for that lies in the structure of the fracton symmetry \eqref{dA}, characterized by two derivatives, which prevents the presence, at the right hand side of \eqref{[DAt,A]} and \eqref{[DA,At]}, of the central charge term $\partial\delta$, typical of usual KM algebras. This leads us to guess that a conserved current algebra might exist on the boundary of LG, whose defining symmetry \eqref{diff} depends on one derivative only. This would be in agreement with the conjecture concerning the existence of a KM algebra in LG mentioned in \cite{Hofman}.

\section{The induced 3D theory}\label{sec-3Dmodel}

\subsection{Canonical variables}

We look for the transformations of the boundary fields $a_{ij}(X),\ \tilde a_{ij}(X)$ which preserve the solutions \eqref{sol1} and \eqref{sol2}. The most general ones are
	\begin{empheq}{align}
	\delta a_{mn}=&\eta_{mn}\phi+\partial_m\xi_n+\partial_n\xi_m+\partial_m\partial_n\lambda\quad;\quad\delta\tilde a_{mn}=0\label{damn}\\
	\tilde\delta\tilde a_{mn}=&\eta_{mn}\tilde\phi+\partial_m\tilde\xi_n+\partial_n\tilde\xi_m+\partial_m\partial_n\tilde\lambda\quad;\quad\tilde\delta a_{mn}=0\ ,\label{damnt}
	\end{empheq}
where $\lambda(X),\ \tilde\lambda(X),\ \phi(X),\ \tilde\phi(X),\ \xi(X)$ and $\tilde\xi(X)$ are generic local parameters.
The solutions $A_{ij}(X)$ \eqref{sol1}  and $\tilde A_{ij}(X)$ \eqref{sol2} remain unchanged, $i.e.$ $\delta A_{ij}=\tilde\delta\tilde A_{ij}=0$, if $\xi_m=\partial_m\xi',\ \tilde\xi_m=\partial_m\tilde\xi'$, so that \eqref{damn} and \eqref{damnt} reduce to
	\begin{empheq}{align}
	\delta a_{mn}=&\eta_{mn}\phi+\partial_m\partial_n\lambda\quad;\quad\delta\tilde a_{mn}=0\label{da1}\\
	\tilde\delta\tilde a_{mn}=&\eta_{mn}\tilde\phi+\partial_m\partial_n\tilde\lambda\quad;\quad\tilde\delta a_{mn}=0\ .\label{dat1}
	\end{empheq}
We decompose the boundary fields $a_{ij}(X)$ and $\tilde a_{ij}(X)$ in terms of their trace and traceless contributions, $i.e.$
	\begin{empheq}{align}
	a_{ij}&=\alpha_{ij}+\frac{1}{3}\eta_{ij}a\\
	\tilde a_{ij}&=\tilde\alpha_{ij}+\frac{1}{3}\eta_{ij}\tilde a\ ,
	\end{empheq}
where $a\equiv\eta^{ij}a_{ij},\ \tilde a\equiv\eta^{ij}\tilde a_{ij}$, and $\alpha_{ij}(X),\ \tilde\alpha_{ij}(X)$ are symmetric traceless fields
\be
\eta^{ij}\alpha_{ij}=\eta^{ij}\tilde\alpha_{ij}=0\ ,
\label{tracelessalpha}\ee
which transform as
	\begin{empheq}{align}
	\delta \alpha_{mn}=&\partial_m\partial_n\lambda-\frac{1}{3}\eta_{mn}\partial^2\lambda\quad;\quad\delta\tilde \alpha_{mn}=0\label{dalpha}\\
	\tilde\delta\tilde \alpha_{mn}=&\partial_m\partial_n\tilde\lambda-\frac{1}{3}\eta_{mn}\partial^2\tilde\lambda\quad;\quad\tilde\delta \alpha_{mn}=0\ .\label{dalphat}
	\end{empheq}
The solutions  \eqref{sol1} and \eqref{sol2} depend only on the traceless components
	\be\label{a,at sol}
	\tilde A^{ij}(X)\equiv\epsilon^{iab}\partial_a \tilde \alpha_b^{\ j}(X)+\epsilon^{jab}\partial_a \tilde \alpha_b^{\ i}(X)\quad;\quad A^{ij}(X)\equiv\epsilon^{iab}\partial_a \alpha_b^{\ j}(X)+\epsilon^{jab}\partial_a \alpha_b^{\ i}(X)\ ,
	\ee	
and the trace contributions disappear. The DoF of the boundary theory are then described by the rank-2 traceless tensor fields $\alpha_{ij}(X)$ and $\tilde\alpha_{ij}(X)$. This is consistent with the fact that, as we showed,  on the boundary the bulk fields $A_{ij}(x)$ and $\tilde A_{ij}(x)$ have only five components, exactly as the 3D boundary fields $\alpha_{ij}(X)$ and $\tilde\alpha_{ij}(X)$, which are symmetric and traceless. The solutions \eqref{a,at sol} highly simplify the definitions of $\Delta A(X)$  \eqref{DA} and $\Delta\tilde A(X)$ \eqref{DAt} :
	\begin{empheq}{align}
	\Delta A|_\eqref{sol2}&=4(g_1-g_2)\epsilon^{0\textsc{mn}}\partial_\textsc{m}\partial^\textsc{a}\alpha_{\textsc{na}}\\
	\Delta \tilde A|_\eqref{sol1}&=4(g_1-g_2)\epsilon^{0\textsc{mn}}\partial_\textsc{m}\partial^\textsc{a}\tilde \alpha_{\textsc{na}}\ ,
	\end{empheq}
where we observe that only spatial derivatives appear. Considering the following combinations of the commutators \eqref{[DAt,A]} and \eqref{[DA,At]} and their traces 
	\begin{empheq}{align}
	\left[\Delta\tilde A\ ,\ A_{\textsc{df}}'-\frac{1}{2}\eta_{\textsc{df}}\eta^{\textsc{mn}}A_{\textsc{mn}}'\right]=&\quad\frac{i}{2}\left(\delta^{\textsc{m}}_{\textsc{d}}\delta^{\textsc{n}}_{\textsc{f}}+\delta^{\textsc{n}}_{\textsc{d}}\delta^{\textsc{m}}_{\textsc{f}}-\eta^{\textsc{mn}}\eta_{\textsc{df}}\right)\partial_\textsc{m}\partial_\textsc{n}\delta^{(2)}(X-X')\label{DAt,A-TrA}\\
	\left[\Delta A\ ,\ \tilde A_{\textsc{df}}'-\frac{1}{2}\eta_{\textsc{df}}\eta^{\textsc{mn}}\tilde A_{\textsc{mn}}'\right]=&-\frac{i}{2}\left(\delta^{\textsc{m}}_{\textsc{d}}\delta^{\textsc{n}}_{\textsc{f}}+\delta^{\textsc{n}}_{\textsc{d}}\delta^{\textsc{m}}_{\textsc{f}}-\eta^{\textsc{mn}}\eta_{\textsc{df}}\right)\partial_\textsc{m}\partial_\textsc{n}\delta^{(2)}(X-X')\ ,
	\end{empheq}
and using the solutions \eqref{a,at sol}, we can  identify the following two canonical commutation relations at the boundary (the details can be found in Appendix \ref{appComm})
	\begin{empheq}{align}
	\left[q_{\textsc{ab}}\ ,\ p'^{\textsc{cd}}\right]=&\frac{i}{2}\left(\delta^{\textsc{c}}_{\textsc{a}}\delta^{\textsc{d}}_{\textsc{b}}+\delta^{\textsc{d}}_{\textsc{a}}\delta^{\textsc{c}}_{\textsc{b}}-\eta^{\textsc{cd}}\eta_{\textsc{ab}}\right)\delta^{(2)}(X-X')\label{al,ft}\\
	\left[\tilde q_{\textsc{ab}}\ ,\ \tilde p'^{\textsc{cd}}\right]=&\frac{i}{2}\left(\delta^{\textsc{c}}_{\textsc{a}}\delta^{\textsc{d}}_{\textsc{b}}+\delta^{\textsc{d}}_{\textsc{a}}\delta^{\textsc{c}}_{\textsc{b}}-\eta^{\textsc{cd}}\eta_{\textsc{ab}}\right)\delta^{(2)}(X-X')\ ,\label{alt,f}
	\end{empheq}
where
	\begin{empheq}{align}
	q_{\textsc{ab}}&\equiv \alpha _{\textsc{ab}}-\frac{1}{2}\eta_{\textsc{ab}} \alpha ^\textsc{m}_{\ \textsc{m}}\\
	p^{\textsc{cd}}&\equiv 2g_{12}\left(\tilde f^{\textsc{cd}0}-\frac{1}{2}\eta^{\textsc{cd}}\tilde f^{\ a0}_{a}\right)\\
	\tilde q_{\textsc{ab}}&\equiv\tilde \alpha _{\textsc{ab}}-\frac{1}{2}\eta_{\textsc{ab}}\tilde \alpha ^\textsc{m}_{\ \textsc{m}}\\
	\tilde p^{\textsc{cd}}&\equiv -2g_{12}\left(f^{\textsc{cd}0}-\frac{1}{2}\eta^{\textsc{cd}}f^{\ a0}_{a}\right)\ ,
	\end{empheq}
and
	\be
	g_{12}\equiv2(g_1-g_2)\ .
	\ee
In analogy to $F_{\mu\nu\rho}(x)$ \eqref{Fmunurho}, $f_{abc}(X)$ and $\tilde f_{abc}(X)$ are defined as
	\begin{empheq}{align}
	\tilde f_{abc}&\equiv\partial_a\tilde \alpha_{bc}+\partial_b\tilde \alpha_{ac}-2\partial_c\tilde \alpha_{ab}\\
	f_{abc}&\equiv\partial_a\alpha_{bc}+\partial_b\alpha_{ac}-2\partial_c\alpha_{ab}\ .
	\end{empheq}	
It is interesting to notice that a canonical commutator  similar to those we found in \eqref{al,ft} and \eqref{alt,f}, appears in \cite{Du:2021pbc}, in the context of the traceless fracton models  \cite{Pretko:2016lgv,Pretko:2016kxt}. 
The aim of \cite{Du:2021pbc} is to build a non-abelian model for fractons in 2+1 dimensions. To do so the abelian traceless theory needs to be defined first. As for any fracton theory \cite{Nandkishore:2018sel,Pretko:2020cko,Pretko:2016lgv,Pretko:2016kxt}, the ``electric field'' $E_\textsc{ij}(x)$ is the conjugate momentum of $A_\textsc{ij}(x)$, from which the commutator holds
	\be
	[E^{\textsc{ij}},A_\textsc{mn}]=i(\delta^\textsc{i}_\textsc{m}\delta^\textsc{j}_\textsc{n}+\delta^\textsc{j}_\textsc{m}\delta^\textsc{i}_\textsc{n})\label{fract-comm}\ .
	\ee
After that, the scalar Gauss constraints is imposed, together with a tracelessness condition : 
	\be\label{Gauss-constr}
	\partial_\textsc{i}\partial_\textsc{j}E^\textsc{ij}=\rho\quad;\quad E^{\ \textsc{i}}_\textsc{i}=0\ ,
	\ee
 which imply  three conservation equations : 
	\be
	\int\rho=const\quad;\quad\int\vec x\rho=const\quad;\quad\int x^2\rho=const\ ,
	\ee
of charge, dipole and a component of the quadrupole, respectively. The main characteristic of fracton theories, $i.e.$ the limited mobility, is here translated to the fact that single charges cannot move, while dipole bound states can only move along their transverse direction. The constraints \eqref{Gauss-constr} imply that the tensor field $A_\textsc{ij}(x)$ transforms exactly as \eqref{da1} and \eqref{dat1}, which is a remarkable check of our reasoning. However, while in our case it is natural to identify the DoF of the theory with the  traceless fields $\alpha_{ij}(X)$ and $\tilde\alpha_{ij}(X)$, in \cite{Du:2021pbc} the tracelessness condition is imposed as a kind of gauge fixing, while for us it comes from the solutions \eqref{a,at sol}. As a consequence, the definition \eqref{fract-comm} of the canonical commutator is no longer valid (since $A^{\ \textsc{i}}_\textsc{i}=E^{\ \textsc{i}}_\textsc{i}=0$ would not commute), and  the commutator for the traceless theory of fractons is defined as Dirac brackets \cite{Dirac:1950pj}, which turns out to be identical to ours \eqref{al,ft} and \eqref{alt,f}.
\normalcolor

\subsection{The most general 3D action}\label{sec S3D}

The action of the 3D boundary theory is constructed as the most general local integrated functional of the traceless rank-2 symmetric tensor fields $\alpha_{ij}(X)$ and $\tilde\alpha_{ij}(X)$ compatible with
	\bi
	\item power-counting $[\alpha]=0,\ [\tilde\alpha]=1$ ;
	\item symmetry $\delta S=\tilde\delta S=0$, where $\delta$ and $\tilde\delta$ are defined in \eqref{dalpha} and \eqref{dalphat} ;
	\item canonical variables identified in \eqref{al,ft} and \eqref{alt,f} :  $\frac{\partial\mathcal L_{kin}}{\partial \dot q}=p$.
	\ei
In Appendix  \ref{appSgen} we show that the most general 3D action satisfying these three requests is
	\be \label{S3D}
	S_{3D}=\int d^3X\left(-\frac{2}{3}g_{12}\,\varphi_{abc}\tilde\varphi^{abc}+\omega_5\, \tilde\alpha^d_a\epsilon^{abc}\tilde\varphi_{dbc}\right)\ ,
	\ee
where we defined
	\be\label{phi}
		\begin{split}
		\varphi_{abc}=\varphi_{bac}&\equiv f_{abc}+\frac{1}{4}\left(-2\eta_{ab}f^d_{\ dc}+\eta_{bc}f^d_{\ da}+\eta_{ac}f^d_{\ db}\right)\\
		&=-2\partial_c\alpha_{ab}+\partial_a\alpha_{bc}+\partial_b\alpha_{ac}-\eta_{ab}\partial^d\alpha_{dc}+\frac{1}{2}\eta_{bc}\partial^d\alpha_{da}+\frac{1}{2}\eta_{ac}\partial^d\alpha_{db}\ ,
		\end{split}
	\ee
and, analogously, $\tilde\varphi_{abc}(X)$ in terms of $\tilde\alpha_{ab}(X)$, with the following properties
	\begin{empheq}{align}
	&\varphi_{abc}+\varphi_{cab}+\varphi_{bca}=0=\tilde\varphi_{abc}+\tilde\varphi_{cab}+\tilde\varphi_{bca}\label{cicl}\\
	&\eta^{ab}\varphi_{abc}=\eta^{bc}\varphi_{abc}=\eta^{ab}\tilde\varphi_{abc}=\eta^{bc}\tilde\varphi_{abc}=0\ .\label{traceless}
	\end{empheq}
The $\omega_5$ term in \eqref{S3D} looks like a CS term, and the similarity is even more evident if we explicit the $\tilde\alpha_{ab}(X)$ dependence, since $ \tilde\alpha^d_a\epsilon^{abc}\tilde\varphi_{dbc}\propto\,\tilde\alpha^d_a\epsilon^{abc}\partial_b\tilde\alpha_{cd}$.
Intriguingly, this CS-like term resembles the massless limit of 3D self-dual massive gravity \cite{Dalmazi:2020xou,Aragone}. This theory contains a 3D Fierz-Pauli mass term that breaks the gauge invariance \cite{Blasi:2017pkk,Blasi:2015lrg}. However, it has been shown \cite{Dalmazi3} that it is dual to linearized topologically massive gravity \cite{Deser}, which is gauge invariant and contains the CS-like term together with the linearized 3D Einstein-Hilbert action. These two equivalent theories were originally proposed as a viable way to describe a single propagating massive graviton in 3D in contrast with the standard Einstein-Hilbert theory, which is topological in 3D and does not support any propagating spin-2 particle. In our case, the linearized Einstein-Hilbert term is replaced by the tensorial Maxwell-like term such that our boundary action still supports a propagating ``graviton''. Notice that a non-covariant version of our CS-like term has been also considered in \cite{Cappelli:2015ocj,Gromov:2017vir} in the context of fractional quantum Hall effect and chiral fractons. However, these non-covariant field theories, that can be seen as dual one to each other, do not take into account any tensorial Maxwell-like terms.
 Notice also that if $\omega_5=0$ it is possible to decouple the fields. In fact, by defining 
	\be
	\alpha_{ab}^\pm\equiv\sqrt{M}\,\alpha_{ab}\pm\frac{1}{\sqrt{M}}\tilde\alpha_{ab}\quad\Rightarrow\quad\varphi^\pm_{abc}=\sqrt{M}\,\varphi_{abc}\pm\frac{1}{\sqrt{M}}\tilde\varphi_{abc}\ ,
	\ee
where $M$ is a parameter with mass dimension $[M]=1$ and $[\alpha^\pm]=\frac{1}{2}$, the 3D action \eqref{S3D} becomes
	\be
	S_{3D}=\int d^3X\left[\frac{g_{12}}{6}\left(\varphi^-_{abc}\varphi^{-\,abc}-\varphi^+_{abc}\varphi^{+\,abc}\right)+\frac{M\omega_5}{4}\left( \alpha^{+\,d}_{\,a}\epsilon^{abc}\varphi^+_{dbc}-2\alpha^{+\,d}_{\,a}\epsilon^{abc}\varphi^-_{dbc}+\alpha^{-\,d}_{\,a}\epsilon^{abc}\varphi^-_{dbc}\right)\right]\ ,
	\ee
which for $\omega_5=0$ decouples:
	\be\label{S3Ddec}
	S_{3D}[\alpha^+,\alpha^-,\omega_5=0]=S^+_{3D}[\alpha^+]+S^-_{3D}[\alpha^-]\
	\ee
with
	\be
	S^\pm_{3D}\equiv\mp\frac{g_{12}}{6}\int d^3X\,\varphi^\pm_{abc}\varphi^{\pm\,abc}\ .
	\ee
As we shall show in Section 7, this second case keeps the $\TR$-invariance of the boundary in agreement with the $\TR$-symmetry of the bulk action (3.1).

\section{The bulk and the boundary: holographic contact}\label{sec-hc}

Once the most general 3D action \eqref{S3D} has been derived, we have to establish the ``holographic contact'' between this induced 3D theory and the 4D theory $S_{tot}$ \eqref{Stot}. This is accomplished by requiring that the EoM of the 3D theory coincide with the BC \eqref{bcA3i}, \eqref{bcAij}, \eqref{bcAtij} of the 4D theory. To do so we have at our disposal the $\xi_i$ parameters appearing in $S_{bd}$ \eqref{Sbd} and $\omega_5$ in $S_{3D}$ \eqref{S3D}. The EoM of $S_{3D}$ are
	\be\label{eom1bd}
		\frac{\delta S_{3D}}{\delta\alpha_{mn}}  
=-2g_{12}\partial_a\tilde\varphi^{mna}=0\ ,
	\ee
where we used the cyclic property \eqref{cicl}, and
	\be\label{eom2bd}
		\frac{\delta S_{3D}}{\delta\tilde\alpha_{mn}}
		=-2g_{12}\partial_a\varphi^{mna}+\omega_5\left(\epsilon^{mab}\tilde\varphi^{n}_{\ ab}+\epsilon^{nab}\tilde\varphi^{m}_{\ ab}\right)=0\ .
	\ee
We now consider the BC of the bulk theory \eqref{bcA3i}, \eqref{bcAij} and \eqref{bcAtij}, which we write in terms of the solutions \eqref{a,at sol} and of the definitions of $\varphi_{abc}(X),\ \tilde\varphi_{abc}(X)$ \eqref{phi} 
	\begin{empheq}{align}
	&{\frac{1}{3}\partial^a\left(\epsilon_{aij}\varphi_b^{\ ij}+\epsilon_{bij}\varphi_a^{\ ij}\right)}=0
\label{bc1-bd}
\\[10px]
	&\frac{2}{3}\xi_0\left(\epsilon^{aij}\varphi^b_{\ ij}+\epsilon^{bij}\varphi^a_{\ ij}\right)+\frac{1}{3}\left(\xi_1-g_{12}\right)\left(\epsilon^{aij}\tilde\varphi^b_{\ ij}+\epsilon^{bij}\tilde\varphi^a_{\ ij}\right)-2\xi_2\partial_c\varphi^{abc}=0\label{bc2-bd}\\[10px]
	&
\frac{\xi_1}{3}\left(\epsilon^{aij}\varphi^b_{\ ij}+\epsilon^{bij}\varphi^a_{\ ij}\right)=0\ .\label{bc3-bd}
	\end{empheq}
The contact is governed by two coefficients : $\xi_1$, which appears in $S_{bd}$ \eqref{Sbd}, and $\omega_5$ in the action $S_{3D}$ \eqref{S3D}. The first - $\xi_1$ - is relevant because it determines the existence of the BC \eqref{bc3-bd}, the second - $\omega_5$ - decouples the EoM of the boundary fields $\alpha_{ab}(X),\ \tilde\alpha_{ab}(X)$, $i.e.$ eliminates the CS-like term from the action  \eqref{S3D}. Additionally, we remark that $\tilde\alpha_{ij}(X)$ appears only in the BC \eqref{bc2-bd}, coupled to $(\xi_1-g_{12})$, which should not vanish, otherwise no contact is possible. To summarize, the constraints on the coefficients, up to now, are
	\be\label{vincoli}
	g_1\neq0\quad;\quad g_{12}\neq0\quad;\quad\xi_1\neq g_{12}\ .
	\ee
Therefore, depending on $\xi_1$ and $\omega_5$, we distinguish the following cases :	
\bi
\item
$\pmb{\xi_1\neq0,\ \omega_5\neq0}$ : using \eqref{bc3-bd} in \eqref{bc2-bd} (or setting $\xi_0=0$), we have
	\be\label{bc2-3}
	\frac{1}{3}\left(\xi_1-g_{12}\right)\left(\epsilon^{aij}\tilde\varphi^b_{\ ij}+\epsilon^{bij}\tilde\varphi^a_{\ ij}\right)-2\xi_2\partial_c\varphi^{abc}=0\ ,
	\ee
which coincides with the 3D EoM \eqref{eom2bd}
	\be
	-2g_{12}\partial_a\varphi^{mna}+\omega_5\left(\epsilon^{mab}\tilde\varphi^{n}_{\ ab}+\epsilon^{nab}\tilde\varphi^{m}_{\ ab}\right)=0
	\ee	
if
	\be\label{hc1}
	\omega_5=\frac{1}{3}(\xi_1-g_{12})\neq0\quad;\quad \xi_2=g_{12}\neq0\quad;\quad\xi_1\neq0\ .
	\ee
Notice that setting $\omega_5=0$ would imply $\xi_1=g_{12}$, which we excluded in  \eqref{vincoli}. Up to a numerical coefficient, we now consider the following symmetric combination of the curl of the BC  \eqref{bc3-bd}
	\be\label{curlBC3}
		\begin{split}
		0&=\epsilon_{mac}\partial^c\eqref{bc3-bd}^a_b+\epsilon_{bac}\partial^c\eqref{bc3-bd}^a_m\\
		&=6\partial^i\varphi_{bmi}\ ,
		\end{split}
	\ee
where we used the properties of tracelessness \eqref{traceless} and cyclicity  \eqref{cicl} of $\varphi_{abc}(X)$. We then use this result in the BC \eqref{bc2-3}, which becomes
	\be\label{bc2-3-d3}
	\frac{1}{3}\left(\xi_1-g_{12}\right)\left(\epsilon^{aij}\tilde\varphi^b_{\ ij}+\epsilon^{bij}\tilde\varphi^a_{\ ij}\right)=0\ ,
	\ee
of which we compute again the curl
	\be
	0=\epsilon_{mac}\partial^c\eqref{bc2-3-d3}^a_b+\epsilon_{bac}\partial^c\eqref{bc2-3-d3}^a_m=2\left(\xi_1-g_{12}\right)\partial^i\tilde\varphi_{bmi}\ ,
	\ee
which finally coincides with the 3D EoM \eqref{eom1bd}
	\be
	-2g_{12}\partial_a\tilde\varphi^{mna}=0\ .
	\ee
Notice that this second contact is obtained without the need of any additional constraint on the parameters, we just need \eqref{hc1}. Taking into account \eqref{hc1}, the 3D action  \eqref{S3D} becomes
	\be \label{S3Dhc}
	S_{3D}=\frac{1}{3}\int d^3X\left[-2g_{12}\,\varphi_{abc}\tilde\varphi^{abc}+(\xi_1-g_{12})\, \tilde\alpha^d_a\epsilon^{abc}\tilde\varphi_{dbc}\right]\ ,
	\ee
while the boundary term \eqref{Sbd} now is
	\be
	S_{bd}=\int d^4x\delta(x^3)\left[{\xi_0} A_{ab}A^{ab}+\xi_1\tilde A_{ab}A^{ab}+g_{12}\epsilon^{abc}A_{ai}\partial_bA_c^i+{\xi_3}(A^a_a)^2+{\xi_4}\tilde A^a_aA^b_b\right]\ ,
	\ee
where the coefficients $\xi_0,\ \xi_3$ and $\xi_4$ are free and can for instance be set to zero, while $\xi_1\neq\{0,g_{12}\}$.
\item
$\pmb{\xi_1\neq0,\ \omega_5=0}$ : the EoM of the 3D theory \eqref{eom1bd} and \eqref{eom2bd} are 
	\be\label{eom-bd-ro=0}
	\partial_a\varphi^{mna}=0\quad;\quad	\partial_a\tilde\varphi^{mna}=0\ ,
	\ee
while, ignoring the first BC \eqref{bc1-bd}, which is automatically solved by the third one \eqref{bc3-bd}, and using \eqref{bc3-bd} in \eqref{bc2-bd}, the remaining BC are
	\begin{empheq}{align}
	\frac{1}{3}\left(\xi_1-g_{12}\right)\left(\epsilon^{aij}\tilde\varphi^b_{\ ij}+\epsilon^{bij}\tilde\varphi^a_{\ ij}\right)-2\xi_2\partial_c\varphi^{abc}\label{bc2-bd'}&=0\\[10px]
	\frac{\xi_1}{3}\left(\epsilon^{aij}\varphi^b_{\ ij}+\epsilon^{bij}\varphi^a_{\ ij}\right)&=0\ .\label{bc3-bd'}
	\end{empheq}
As in \eqref{curlBC3}, we can again compute 
	\be\label{rotBC2}
	0=\epsilon_{mac}\partial^c\eqref{bc3-bd'}^a_b+\epsilon_{bac}\partial^c\eqref{bc3-bd'}^a_m=2\xi_1\partial^i\varphi_{bmi}
	\ee
which coincides with the first EoM of \eqref{eom-bd-ro=0}. If we use this result \eqref{rotBC2} in \eqref{bc2-bd'} (analogous to setting $\xi_2=0$) and consider the same combination as \eqref{rotBC2}, we obtain the second EoM of \eqref{eom-bd-ro=0} and thus we get the second matching between bulk and boundary. In that case the 3D action \eqref{S3D} is
	\be \label{S3Dhc}
	S_{3D}=-\frac{2g_{12}}{3}\int d^3X\,\varphi_{abc}\tilde\varphi^{abc}\ ,
	\ee
while the boundary term $S_{bd}$ \eqref{Sbd} becomes 
	\be
	S_{bd}=\int d^4x\delta(x^3)\left[{\xi_0} A_{ab}A^{ab}+\xi_1\tilde A_{ab}A^{ab}+{\xi_2}\epsilon^{abc}A_{ai}\partial_bA_c^i+{\xi_3}(A^a_a)^2+{\xi_4}\tilde A^a_aA^b_b\right]\ ,
	\ee
where the coefficients $\xi_0,\ \xi_2,\ \xi_3$ and $\xi_4$ are free, $i.e.$ do not contribute to the contact between the bulk and the boundary and can be set to zero without loss of generality, provided that  $\xi_1\neq\{0,g_{12}\}$. Therefore a second holographic contact is possible if
	\be\label{hc2}
	\omega_5=0\quad;\quad\xi_1\neq\{0,g_{12}\}\ .
	\ee
This second result has a relevant consequence : $\omega_5=0$ allows to decouple the action, as seen  in \eqref{S3Ddec}.
\ei
We observe that the holographic contacts obtained in \eqref{hc1} and \eqref{hc2} affect the boundary action $S_{bd}$ \eqref{Sbd} in different ways, in particular in the first case \eqref{hc1} the number of free parameters from five reduces to three, while in the second case \eqref{hc2} it goes to four. Finally, if $\xi_1=0$, no complete matching between BC and 3D EoM is possible, in fact setting ${\xi_1=0}$, one of the BC \eqref{bc3-bd} disappears. We are left with
	\begin{empheq}{align}
	&{\frac{1}{3}\partial^a\left(\epsilon_{aij}\varphi_b^{\ ij}+\epsilon_{bij}\varphi_a^{\ ij}\right)}=0
\label{bc1-bd''}\\[10px]
	&\frac{2}{3}\xi_0\left(\epsilon^{aij}\varphi^b_{\ ij}+\epsilon^{bij}\varphi^a_{\ ij}\right)-\frac{g_{12}}{3}\left(\epsilon^{aij}\tilde\varphi^b_{\ ij}+\epsilon^{bij}\tilde\varphi^a_{\ ij}\right)-2\xi_2\partial_c\varphi^{abc}=0\label{bc2-bd''}\ .
	\end{empheq}
For what concerns the parameter $\omega_5$ in \eqref{S3D}, two cases are possible
\bi
\item
$\pmb{\omega_5\neq0}$ : the BC \eqref{bc2-bd''} coincides with the 3D EoM \eqref{eom2bd} 
	\be
	-2g_{12}\partial_a\varphi^{mna}+\omega_5\left(\epsilon^{mab}\tilde\varphi^{n}_{\ ab}+\epsilon^{nab}\tilde\varphi^{m}_{\ ab}\right)=0
	\ee	
if
	\be\label{hc3}
	\omega_5=-\frac{g_{12}}{3}\neq0\quad;\quad \xi_2=g_{12}\neq0\quad;\quad\xi_0=\xi_1=0\ ,
	\ee
however it is not possible to establish a link with the other EoM \eqref{eom1bd}.
\item
$\pmb{\omega_5=0}$ : the EoM of the 3D boundary theory are given by \eqref{eom-bd-ro=0}. A matching is possible with the BC \eqref{bc2-bd''} only if $\xi_2=0$, and if we compute
	\be
	0=\epsilon_{mac}\partial^c\eqref{bc2-bd''}^a_b+\epsilon_{bac}\partial^c\eqref{bc2-bd''}^a_m=4\xi_0\partial^i\varphi_{bmi}-2g_{12}\partial^i\tilde\varphi_{bmi}\ ,
	\ee
which coincides with  a combination of the two EoM of the boundary. We see that also in this case a complete holographic contact between 3D EoM and bulk BC, is not possible.
\ei
This enforces the fact that the $\xi_1$-term in $S_{bd}$ \eqref{Sbd} plays a key role in the holographic contact. To summarize
\begin{table}[H]
\centering
  \begin{tabular}{ | c| c |  }
    \hline
     & BC-EoM matching \\ \hline
      $\xi_1\neq0,\ \omega_5\neq0$ & $\omega_5=\frac{1}{3}(\xi_1-g_{12})\ ;\ \xi_2=g_{12}$\\ \hline
      $\xi_1\neq0,\ \omega_5=0$ & $\omega_5=0$  \\ \hline
      $\xi_1=0$ & no contact \\ \hline
  \end{tabular}
  \caption{\footnotesize\label{table2} Holographic contacts.}
  \end{table}

\section{Physical interpretation of the 3D theory}\label{sec-phys}

To understand the physical content of the 3D theory described by the action $S_{3D}$ \eqref{S3D}, we study its EoM. The first EoM \eqref{eom1bd} for $m=n=0$ gives
	\be\label{eom1bd00}
	0=\partial_{\textsc{a}}\tilde\varphi^{00\textsc{a}}=-2\partial_{\textsc{a}}\partial^\textsc{a}\tilde\alpha^{00}+2\partial_\textsc{a}\partial^0\tilde\alpha^{\textsc{a}0}+\partial_\textsc{a}\partial_b\tilde\alpha^{\textsc{a}b}	\ .
	\ee
Taking the $\partial_\textsc{n}$-derivative of \eqref{eom1bd} for $m=0,\ n=\textsc n$, we get
	\be\label{Gauss}
		0=\partial_\textsc{n}\partial_a\tilde\varphi^{0\textsc{n}a}
		=-\frac{1}{4g_{12}}\partial_\textsc{a}\partial_\textsc{n}p^{\textsc{a}\textsc{n}}\ ,
	\ee
where we used \eqref{eom1bd00}, the cyclicity property \eqref{cicl} and the definition of conjugate momentum in terms of $\tilde\varphi_{abc}(X)$ \eqref{phi}
	\be
 	p^{\textsc{mn}}=2g_{12}\tilde\varphi^{\textsc{mn}0}\label{p}\ .
	\ee
We see that \eqref{Gauss} is a Gauss-like equation, analogous to the one related to the traceless scalar charge model of fractons in vacuum \cite{Pretko:2016lgv,Pretko:2016kxt}, which is at the base of the limited mobility property. We therefore realize that the induced 3D theory shows fractonic properties. To analyze the second EoM \eqref{eom2bd}, we first compute the conjugate momentum of $\tilde\alpha_{ab}(X)$ :
	\be\label{pt}
	\tilde p^{\textsc{mn}}=\frac{\partial \mathcal L_{3D}}{\partial\dot{\tilde\alpha}_{\textsc{mn}}}
		=2g_{12}\varphi^{\textsc{mn}0}-\frac{3}{2}\omega_5\left(\epsilon^{0\textsc{am}}\tilde\alpha^\textsc{n}_{\ \textsc{a}}+\epsilon^{0\textsc{an}}\tilde\alpha^\textsc{m}_{\ \textsc{a}}\right)\ .
	\ee
The EoM for $\tilde\alpha_{ab}(X)$ \eqref{eom2bd} at $m=n=0$ is
		\be\label{eom2bd00}
		\partial_\textsc{a}\varphi^{00\textsc{a}}=\frac{3\omega_5}{g_{12}}\epsilon^{0\textsc{ab}}\partial_\textsc{a}\tilde\alpha^0_{\ \textsc{b}}\ ,
		\ee
and, as in the previous case, taking the $\partial_\textsc{n}$-derivative of \eqref{eom2bd} at $m=0,\ n=\textsc n$ we have
		\be
			0=-2g_{12}\partial_a\partial_\textsc{n}\varphi^{0\textsc{n}a}+3\omega_5\partial_\textsc{n}\left(\epsilon^{0\textsc{ab}}\partial_\textsc{a}\tilde\alpha^\textsc{n}_{\ \textsc{b}}+\epsilon^{\textsc{n}ab}\partial_a\tilde\alpha^0_{\ b}\right)
			=\frac{1}{2}\partial_\textsc{m}\partial_\textsc{n}\tilde p^{\textsc{mn}}-\frac{3}{2}\omega_5\epsilon^{0\textsc{am}}\partial_\textsc{m}\partial_\textsc{n}\tilde\alpha^\textsc{n}_{\ \textsc{a}}\ ,
		\ee
where we used the cyclic property of $\varphi_{abc}(X)$ \eqref{cicl} and \eqref{eom2bd00}. Here again we find a Gauss-like equation for the traceless scalar charge theory of fractons \cite{Pretko:2016lgv,Pretko:2016kxt}, but with a matter contribution at the right hand side
		\be\label{tGauss6.7}
		\partial_\textsc{m}\partial_\textsc{n}\tilde p^{\textsc{mn}}=\tilde\rho_5=\omega_5\tilde\rho\ ,
		\ee
where
		\be\label{rho}
		\tilde\rho\equiv{3}\epsilon^{0\textsc{am}}\partial_\textsc{m}\partial_\textsc{n}\tilde\alpha^\textsc{n}_{\ \textsc{a}}
		\ee
plays the role of charge. This gives an interesting interpretation of the CS-like term in the induced 3D action \eqref{S3D} as ``internal'' matter. Notice that this term coincides with the charge identified by Pretko in \cite{Pretko:2017xar}, where a non-covariant CS-like term is studied. In that case the CS term comes from a non-covariant fractonic $\theta$-term in the bulk, and it is written in terms of a spatial traceless tensor. The charge $\tilde\rho(X)$ \eqref{rho} comes from a constraint generated by a Lagrange multiplier that is inherited by the CS action from the definition of the $\theta$-term. We observe that this $\tilde\rho(X)$ charge implies, by definition, a dipole conservation. The 3D theory \eqref{S3D} depends on  two fields $\alpha_{ij}(X)$ and $\tilde\alpha_{ij}(X)$, hence the conjugate momenta are two as well  \eqref{p} and \eqref{pt}, which in fracton models play the role of ``electric'' fields :
	\begin{empheq}{align}
	 E^{\textsc{mn}}&\equiv p^{\textsc{mn}}=2g_{12}\tilde\varphi^{\textsc{mn}0}\label{E}\\
	\tilde E^{\textsc{mn}}&\equiv \tilde p^{\textsc{mn}}=2g_{12}\varphi^{\textsc{mn}0}-\frac{3}{2}\omega_5\left(\epsilon^{0\textsc{am}}\tilde\alpha^\textsc{n}_{\ \textsc{a}}+\epsilon^{0\textsc{an}}\tilde\alpha^\textsc{m}_{\ \textsc{a}}\right)\label{tE}\ ,
	\end{empheq}
which satisfy the Gauss equations \eqref{Gauss} and \eqref{tGauss6.7}, which we write
	\begin{empheq}{align}
	\partial_\textsc{m}\partial_\textsc{n}E^{\textsc{mn}}=&0\\
	\partial_\textsc{m}\partial_\textsc{n}\tilde E^{\textsc{mn}}=&\tilde\rho_5\ .\label{tGauss}
	\end{empheq}
Concerning the corresponding ``magnetic'' fields for this theory, inspired by the ordinary 4D electromagnetism, where $B_i\propto\epsilon_{0ijk}F^{jk}$, it is natural to define
	\begin{empheq}{align}
	&B^\textsc{m}\equiv g\epsilon_{0\textsc{ab}}\tilde\varphi^{\textsc{mab}}\quad;\quad	B_\textsc{m}\equiv-g\epsilon^{0\textsc{ab}}\tilde\varphi_{\textsc{mab}}\label{B1}\\
	&\tilde B^\textsc{m}\equiv \tilde g\epsilon_{0\textsc{ab}}\varphi^{\textsc{mab}}\quad;\quad\tilde	B_\textsc{m}\equiv-\tilde g\epsilon^{0\textsc{ab}}\varphi_{\textsc{mab}}\ .\label{Bt1}
	\end{empheq}
Notice that, while in 4D fracton theories both electric and magnetic fields are rank-2 tensors \cite{Bertolini:2022ijb}, in our 3D case, the electric field is still a tensor, while the magnetic field is a vector. Moreover, the definitions \eqref{B1} and \eqref{Bt1} are consistent with the fact that in ordinary 3D electromagnetism the electric field $\vec E(x)$ is a vector, while the magnetic field is a pseudo-scalar $B=\epsilon^{0ij}F_{ij}$ \cite{Boito:2018rdh}. As we shall see, our guess \eqref{B1} and \eqref{Bt1} will be confirmed by a consistent physical interpretation of a fractonic ``magnetic-like'' behaviour. In terms of $\alpha_{ab}(X),\ \tilde\alpha_{ab}(X)$ the magnetic fields read
	\be
	B^\textsc{m}=-3g\epsilon_{0\textsc{ab}}\left(\partial^\textsc{b}\tilde\alpha^\textsc{ma}+\frac{1}{2}\eta^{\textsc{ma}}\partial_d\tilde\alpha^{\textsc b d}\right)\quad;\quad\tilde	B^\textsc{m}=-3\tilde g\epsilon_{0\textsc{ab}}\left(\partial^\textsc{b}\alpha^\textsc{ma}+\frac{1}{2}\eta^{\textsc{ma}}\partial_d\alpha^{\textsc b d}\right)\ ,
	\ee
which imply
	\begin{empheq}{align}
	\tilde\varphi^{\textsc{abc}}&=-\frac{1}{3g}\left(\epsilon^{0\textsc{ac}}B^\textsc{b}+\epsilon^{0\textsc{bc}}B^\textsc{a}\right)\label{B}\\
	\varphi^{\textsc{abc}}&=-\frac{1}{3\tilde g}\left(\epsilon^{0\textsc{ac}}\tilde B^\textsc{b}+\epsilon^{0\textsc{bc}}\tilde B^\textsc{a}\right)\ \label{tB}\ .
	\end{empheq}
Due to tracelessness property \eqref{traceless}, we get 
\be\label{TrPhi=B}
\varphi_{\textsc a}^{\ \textsc{ab}}=\varphi^{00\textsc b}=-\frac{2}{3\tilde g}\epsilon^{0\textsc{ab}}\tilde B_\textsc{a}\quad;\quad \tilde \varphi_{\textsc a}^{\ \textsc{ab}}=\tilde \varphi^{00\textsc b}=-\frac{2}{3g}\epsilon^{0\textsc{ab}}B_\textsc{a}\ .
\ee
Notice that
	\begin{empheq}{align}
	\partial_\textsc{m}B^\textsc{m}=&-\frac{3}{2}g\,\epsilon_{0\textsc{ab}}\partial^\textsc{a}\left(\partial_0\tilde\alpha^{0\textsc b}-\partial_\textsc{c}\tilde\alpha^{\textsc{bc}}\right)\neq0\label{divBneq0}\\
	 \partial_\textsc{m}\tilde B^\textsc{m}=&-\frac{3}{2}\tilde g\,\epsilon_{0\textsc{ab}}\partial^\textsc{a}\left(\partial_0\alpha^{0\textsc b}-\partial_\textsc{c}\alpha^{\textsc{bc}}\right)\neq0\ ,\label{divtBneq0}
	\end{empheq}
which would suggest the presence, in the 3D theory \eqref{S3D}, of a fractonic ``magnetic''-like vortex. Consistently with the fact of having non-vanishing divergences of the magnetic vector fields, we find also a broken Bianchi identity, which also suggests the presence of a kind of magnetic fracton vortex. This would imply that a part of our fracton fields give rise to 2D fracton vortex defects that represent a lower-dimensional version of  the 3D fracton magnetic monopole proposed in \cite{Pretko:2017xar}. In fact, we have
	\be
	\epsilon^{mbc}\partial_m\varphi_{abc}=-\frac{3}{2}\epsilon_{mac}\partial^m\partial_d\alpha^{cd}\neq0\quad;\quad\epsilon^{mbc}\partial_m\tilde\varphi_{abc}=-\frac{3}{2}\epsilon_{mac}\partial^m\partial_d\tilde\alpha^{cd}\neq0 \ ,
	\ee
which for $a=0$ give the non-vanishing divergences \eqref{divBneq0} and \eqref{divtBneq0}. Setting instead $a=\textsc a$  we find
\begin{empheq}{align}
\frac{1}{g}\partial_0B_\textsc{a}+\frac{1}{2g_{12}}\epsilon^{0\textsc{mb}}\partial_\textsc{m}E_{\textsc{ab}}&=\frac{3}{2}\epsilon_{m\textsc a b}\partial^m\partial_d\tilde\alpha^{bd}+\epsilon^{\textsc{m0b}}\partial_\textsc{m}\tilde\varphi_{\textsc{0ab}}\neq0\\
\frac{1}{\tilde g}\partial_0\tilde B_\textsc{a}+\frac{1}{2g_{12}}\epsilon^{0\textsc{mb}}\partial_\textsc{m}\tilde E_{\textsc{ab}}&=\frac{9}{4}\frac{\omega_5}{g_{12}}\partial^\textsc{m}\tilde\alpha_\textsc{ma}+\frac{3}{2}\epsilon_{m\textsc a b}\partial^m\partial_d\alpha^{bd}+\epsilon^{\textsc{m0b}}\partial_\textsc{m}\varphi_{\textsc{0ab}}\neq0\ ,
\end{empheq}
which have non-vanishing right hand sides. Nonetheless, we have two scalar identities for $\varphi_{abc}(X)$ and $\tilde\varphi_{abc}(X)$ :
	\begin{empheq}{align}
	\epsilon^{mbc}\partial_m\partial^a\varphi_{abc}=&-\frac{3}{2}\epsilon_{mac}\partial^m\partial^a\partial_d\alpha^{cd}=0 \\
	\epsilon^{mbc}\partial_m\partial^a\tilde\varphi_{abc}=&-\frac{3}{2}\epsilon_{mac}\partial^m\partial^a\partial_d\tilde\alpha^{cd}=0 \ .
	\end{empheq}
Going back to the 3D EoM, we consider  \eqref{eom1bd} at $m=\textsc m,\ n=\textsc n$
	\be
		\partial_0\tilde\varphi^{\textsc{mn}0}+\partial_\textsc{a}\tilde\varphi^{\textsc{mna}}
		=\frac{1}{2g_{12}}\partial_tE^{\textsc{mn}}-\frac{1}{3g}\partial_\textsc{a}\left(\epsilon^{0\textsc{ma}}B^{\textsc n}+\epsilon^{0\textsc{na}}B^{\textsc m}\right)\ ,
	\ee
where we used the definitions  \eqref{E}, \eqref{B}. We thus have
	\be\label{amp1}
	\partial_tE^{\textsc{mn}}-\frac{2g_{12}}{3g}\partial_\textsc{a}\left(\epsilon^{0\textsc{ma}}B^{\textsc n}+\epsilon^{0\textsc{na}}B^{\textsc m}\right)=0\ ,
	\ee
which, remarkably, coincides with the traceless analog of the Amp\`ere equation of 3D fractons identified in Eq.(16) of \cite{Pretko:2017kvd}, where our \eqref{amp1} is obtained as an EoM  from a 3D Maxwell-like Hamiltonian defined {\it ad hoc}. Differently from our case, the 3D fracton theory in \cite{Pretko:2017kvd} is  not traceless and, in particular,  $E^{\ \textsc{a}}_\textsc{a}\neq0$. The aim of \cite{Pretko:2017kvd} is to study the so called fracton-elasticity duality and, more specifically, the analog of our \eqref{amp1} is used to investigate the effect of creation of defects as consequence of longitudinal motion of dipoles, which in the traceless fracton theory is not present, since dipoles only move along their transverse direction. We keep considering the EoM \eqref{eom2bd} at $m=\textsc m,\ n=\textsc n$ :
	\be
		\begin{split}
		0=&-2g_{12}\partial_a\varphi^{\textsc{mn}a}+\omega_5\left(\epsilon^{\textsc{m}ab}\tilde\varphi^{\textsc{n}}_{\ ab}+\epsilon^{\textsc{n}ab}\tilde\varphi^{\textsc{m}}_{\ ab}\right)\\
		=&-\partial_0\tilde E^{\textsc{mn}}+\frac{3}{2}\omega_5\partial_0\left(\epsilon^{0\textsc{bm}}\tilde\alpha^\textsc{n}_\textsc{b}+\epsilon^{0\textsc{bn}}\tilde\alpha^\textsc{m}_\textsc{b}\right)+\frac{2g_{12}}{3\tilde g}\partial_\textsc{a}\left(\epsilon^{0\textsc{ma}}\tilde B^{\textsc n}+\epsilon^{0\textsc{na}}\tilde B^{\textsc m}\right)+3\omega_5\partial_{\textsc{b}}\left(\epsilon^{0\textsc{mb}}\tilde\alpha^\textsc{n}_0+\epsilon^{0\textsc{nb}}\tilde\alpha^\textsc{m}_0\right)\ ,
		\end{split}
	\ee
where we used the definitions  \eqref{tE}, \eqref{tB}. We have
	\be\label{amp2}
	\partial_t\tilde E^{\textsc{mn}}-\frac{2g_{12}}{3\tilde g}\partial_\textsc{a}\left(\epsilon^{0\textsc{ma}}\tilde B^{\textsc n}+\epsilon^{0\textsc{na}}\tilde B^{\textsc m}\right)=\mathcal{\tilde{J}}_5^{\textsc{mn}}\equiv\omega_5\mathcal{\tilde{J}}^{\textsc{mn}}\ ,
	\ee
which is analogous to the Amp\`ere equation, in presence of a tensorial current, again related to the CS-like term in \eqref{S3D}, which behaves as a matter term
	\be\label{J}
		\mathcal{\tilde{J}}^{\textsc{mn}}
		\equiv3\left[\frac{1}{2}\partial_0\left(\epsilon^{0\textsc{bm}}\tilde\alpha^\textsc{n}_\textsc{b}+\epsilon^{0\textsc{bn}}\tilde\alpha^\textsc{m}_\textsc{b}\right)+\partial_{\textsc{b}}\left(\epsilon^{0\textsc{mb}}\tilde\alpha^\textsc{n}_0+\epsilon^{0\textsc{nb}}\tilde\alpha^\textsc{m}_0\right)\right]\ .
	\ee
By computing  $\partial_\textsc{m}\partial_\textsc{n}$ of \eqref{amp2}, we also get
	\be\label{cont1}
		\partial_t\tilde\rho_5=\partial_\textsc{m}\partial_\textsc{n}\mathcal{\tilde{J}}_5^{\textsc{mn}}\ ,
	\ee
where we used the Gauss-like equation \eqref{tGauss}. Eq.\eqref{cont1} represents a continuity equation typical of scalar fracton theories \cite{Pretko:2016lgv,Pretko:2016kxt}  if $\omega_5\neq0$
	\be\label{cont}
	\partial_t\tilde\rho-\partial_\textsc{m}\partial_\textsc{n}\mathcal{\tilde{J}}^{\textsc{mn}}=0\ .
	\ee
In $\tilde{\mathcal{J}}^{\textsc{mn}}(X)$ \eqref{J}, the contribution associated to the time derivative coincides with the one defined by Pretko (Eq.(118) of \cite{Pretko:2017xar}) as ``generalized Hall response''. As in our case, it is derived from a CS-like term seen as a boundary contribution originated by a fractonic $\theta$-term in the bulk. In particular, it comes from the dynamical part of the action. From the action $S_{3D}$ \eqref{S3D} we can identify both the current and the Amp\`ere-like equation \eqref{amp2}, to which the current \eqref{J} contributes. Moreover, as already mentioned for \eqref{amp1}, this second equation \eqref{amp2} is compatible with the traceful version identified in \cite{Pretko:2017kvd} in the context of an analysis of 3D fracton-elasticity duality. Since $E^{\ \textsc{m}}_\textsc{m}=\tilde E^{\ \textsc{m}}_\textsc{m}=0$, by computing the trace of the Amp\`ere-like equations \eqref{amp1} and \eqref{amp2}, we find
	\begin{empheq}{align}
	\epsilon^{0\textsc{mn}}\partial_\textsc{m}B_\textsc{n}&=0\\
	\epsilon^{0\textsc{mn}}\partial_\textsc{m}\tilde B_\textsc{n}&=-\frac{3}{4}\frac{\tilde g}{g_{12}}\mathcal{\tilde J}^{\textsc{m}}_{5\ \textsc m}=-\frac{9}{2}\frac{g\omega_5}{g_{12}}\epsilon^{0\textsc{mn}}\partial_\textsc{m}\tilde\alpha_{0\textsc n}\ ,
	\end{empheq}
which consistently coincide with  the EoM for $m=n=0$ \eqref{eom1bd00}, \eqref{eom2bd00} previously found, $i.e.$
	\be
	\partial_{\textsc{a}}\tilde\varphi^{00\textsc{a}}=0\quad;\quad\partial_\textsc{a}\varphi^{00\textsc{a}}=\frac{3\omega_5}{g_{12}}\epsilon^{0\textsc{ab}}\partial_\textsc{a}\tilde\alpha^0_{\ \textsc{b}}\ ,
	\ee
due to \eqref{TrPhi=B}. The EoM of the 3D boundary theory may be interpreted as a traceless tensorial extension of the standard 3D Maxwell equations \cite{Boito:2018rdh}, as summarized in Table~3
\begin{table}[H]
\centering
  \begin{tabular}{ | rl | c | c| }
    \hline
   &  &\bf Maxwell &\bf Boundary of LG and fractons \\ \hline
fields& electric, magnetic & $\vec E\ ,\ B$& 	$E^{\textsc{ab}},\ B^\textsc{a}\ ;\ \tilde E^{\textsc{ab}},\ \tilde B^\textsc{a}$ \\ \hline
&in vacuum& $\vec\nabla\cdot\vec E=0$ & $\partial_\textsc{a}\partial_\textsc{b}E^\textsc{ab}=0$ \\
\bf Gauss && $$ & $$ \\

 &with matter& $\vec\nabla\cdot\vec E=\rho$ & $\partial_\textsc{a}\partial_\textsc{b}\tilde E^\textsc{ab}=\tilde\rho_5$ \\ \hline

 &in vacuum& $\partial_t\vec E-\vec\nabla_\bot B=0$ & $	\partial_tE^{\textsc{mn}}-\frac{2g_{12}}{3g}\partial_\textsc{a}\left(\epsilon^{0\textsc{ma}}B^{\textsc n}+\epsilon^{0\textsc{na}}B^{\textsc m}\right)=0$ \\ 

\bf Amp\`ere&&&\\
&with matter&$\partial_t\vec E-\vec\nabla_\bot B=\vec J$ & $\partial_t\tilde E^{\textsc{mn}}-\frac{2g_{12}}{3\tilde g}\partial_\textsc{a}\left(\epsilon^{0\textsc{ma}}\tilde B^{\textsc n}+\epsilon^{0\textsc{na}}\tilde B^{\textsc m}\right)=\mathcal{\tilde{J}}_5^{\textsc{mn}}$ \\ \hline
  \end{tabular}
  \caption{\footnotesize\label{table3} Comparison between EoM and 3D Maxwell.}
\end{table}
where $\vec\nabla_\bot B\equiv\epsilon^\textsc{0ij}\partial_\textsc{j}B$, and the results are consistent with what can be found in the fracton literature \cite{Pretko:2017kvd,Pretko:2016lgv,Pretko:2016kxt,Pretko:2017xar}. We thus recovered, as EoM, the Gauss constraints related to the mobility  of the traceless fracton theory in 3D \cite{Pretko:2016lgv,Pretko:2016kxt}, where the CS-like term contributes as a matter term through $\tilde\rho(X)$ \eqref{rho}, also identified by Pretko in \cite{Pretko:2017xar}. This CS-like term plays the role of matter contribution also in the fractonic Amp\`ere equation \eqref{amp2}, as a current  $\tilde{\mathcal{J}}^{\textsc{mn}}(X)$ \eqref{J}. Here again the term is in accordance with the literature, and in particular with what has been defined as ``generalized Hall response'' in \cite{Pretko:2017xar}. The Amp\`ere equations  \eqref{amp1} and \eqref{amp2}, to which the current $\tilde{\mathcal{J}}^{\textsc{mn}}(X)$ belongs, can be traced back to fracton theories as well, and, more specifically, they have the same structure as the fractonic Amp\`ere equation used in  \cite{Pretko:2017kvd} to study a duality between the theory of fractons and the theory of elasticity. However the one considered in \cite{Pretko:2017kvd} refers to the traceful theory of fractons, whose aim is to study defects as consequence of longitudinal motion of dipoles, which in the traceless fracton theory is not present since dipoles only move along their transverse direction \cite{Pretko:2016lgv,Pretko:2016kxt}. Therefore it would be interesting  to understand if a fracton-elasticity duality also exists for the traceless model. We also notice that in  \cite{Gromov:2017vir} a charge $\tilde\rho(x)$, a current  $\tilde J_\textsc{ab}(x)$, and a continuity equation as \eqref{cont}  are identified  from a CS-like theory with torsion $T(x)$, $i.e. $  $\int d^3x\epsilon^{\mu\nu\rho} e_\mu^\textsc{a}T_{\nu\rho}^\textsc{a}$. In particular, the model coincides with the one proposed in \cite{Pretko:2017xar} for the non-covariant CS-like action for a specific choice of vielbein $e_\mu^\textsc{a}(x)$, and under the condition of ``area-preserving diffeomorphisms'', which seems to be strictly related to fracton models, as also studied in \cite{Du:2021pbc}. This intriguing role of torsion in 3D chiral fractons has been recently extended to 4D fractons \cite{Bertolini:2022ijb}, where it has been shown that a linearized topological term with torsion \cite{Chatzistavrakidis:2020wum} gives rise to the fracton $\theta$-term \cite{Pretko:2017xar}.

\section{Discrete symmetries: parity and time reversal}\label{sec PT}

As extensively shown in the recent literature concerning non-perturbative aspects of quantum field theories, discrete symmetries play a central role in the identification of global anomalies and anomaly inflow, which are related to topological obstructions and impose strong constraints on the renormalization group flows, massive boundary states, quantum dualities and the vacua of quantum field theories \cite{Cordova1,Gaiotto1,Benini,Gaiotto2,Kapustin,Walker,Cordova2}. Moreover, the anomaly inflow has been also extended to certain non-covariant fracton models \cite{Burnell}. Thus, here we analyze some discrete symmetries, such as $\TR$ and $\Par$ in the context of the induced theory derived in the previous Sections.
In fact, we can further constrain the induced 3D action by requiring a matching between the discrete symmetries in the bulk and on the boundary. In particular, under $\Par$ and $\TR$ the bulk fields transform as follows
	\begin{empheq}{align}
	&\TR\left\{A_{00}\,,\,A_{0\textsc{a}}\,,\,A_{\textsc{ab}}\right\}=\left\{A_{00}\,,\,-A_{0\textsc{a}}\,,\,A_{\textsc{ab}}\right\}\quad;\quad\TR\left\{\tilde A_{00}\,,\,\tilde A_{0\textsc{a}}\,,\,\tilde A_{\textsc{ab}}\right\}=\left\{\tilde A_{00}\,,\,-\tilde A_{0\textsc{a}}\,,\,\tilde A_{\textsc{ab}}\right\}\\
	&\Par \left\{A_{00}\,,\,A_{0\textsc{a}}\,,\,A_{\textsc{ab}}\right\}=\left\{A_{00}\,,\,-A_{0\textsc{a}}\,,\,A_{\textsc{ab}}\right\}\quad;\quad\Par \left\{\tilde A_{00}\,,\,\tilde A_{0\textsc{a}}\,,\,\tilde A_{\textsc{ab}}\right\}=\left\{\!-\tilde A_{00}\,,\tilde A_{0\textsc{a}}\,,-\tilde A_{\textsc{ab}}\right\}\\
	&\TR A=A\quad;\quad\TR\tilde A=\tilde A\quad;\quad \Par A=A\quad;\quad \Par \tilde A=-\tilde A
	\ .
	\end{empheq}
The bulk action \eqref{Sbulk} is invariant under $\TR$. Instead, due to the presence of the boundary $x^3=0$, the action is no longer $\Par$-invariant. We now consider the boundary term \eqref{Sbd}, and distinguish between space and time indices
	\be\label{SbdPT}
		\begin{split}
		S_{bd}=&\int d^4x\delta(x^3)\left[\xi_0 \left(A_{00}A^{00}+2A_{0\textsc{a}}A^{0\textsc{a}}+A_{\textsc{ab}}A^{\textsc{ab}}\right)+\xi_1\left(\tilde A_{00}A^{00}+2\tilde A_{0\textsc{a}}A^{0\textsc{a}}+\tilde A_{\textsc{ab}}A^{\textsc{ab}}\right)\right.+\\
		&\qquad\qquad\quad\left.+\xi_2\epsilon^{0\textsc{ab}}\left(A_{0i}\partial_\textsc{a}A_\textsc{b}^i-A_{\textsc{a}i}\partial_0A_\textsc{b}^i+A_{\textsc{a}i}\partial_\textsc{b}A_0^i\right)+\xi_3A^2+\xi_4\tilde AA\right]\ .
		\end{split}
	\ee
	We then observe that
	\bi
	\item $\Par S_{bd}=S_{bd}$ if $\xi_1=\xi_4=0$ ;
	\item $\TR S_{bd}=S_{bd}$ if $\xi_2=0$ ;
	\item $\TR\Par S_{bd}=S_{bd}$ if $\xi_1=\xi_2=\xi_4=0$ .
	\ei
Under these considerations, we can update Table \ref{table2} of holographic contacts with the discrete symmetries allowed on the boundary term $S_{bd}$
	\begin{table}[H]
\centering
  \begin{tabular}{ | l | c | c| }
    \hline
     &BC-EoM matching& Discrete symmetries of $S_{bd}$\\ \hline
      $\xi_1\neq0,\ \omega_5\neq0$ & $\omega_5=\frac{1}{3}(\xi_1-g_{12})\ ;\ \xi_2=g_{12}$& No\\ \hline
      $\xi_1\neq0,\ \omega_5=0$ & $\omega_5=0$ & $\TR$ \\ \hline
  \end{tabular}
  \caption{\footnotesize\label{table4} Holographic contacts, constraints and possible symmetries on $S_{bd}$ \eqref{Sbd}.
}
\end{table}
From Table \ref{table4} we see that imposing $\Par$ on $S_{bd}$ does not lead to a holographic contact, since the $\xi_1$ term, crucial for the existence of the induced 3D action, is not $\Par$-invariant. We also highlight a relation  between $\omega_5\leftrightarrow\xi_2\leftrightarrow\TR$, in fact $\TR$ symmetry is possible only when $\xi_2=0$, which is allowed only in the second holographic contact \eqref{hc2}, $i.e.$ when $\omega_5$ is set to zero as well (in the first case \eqref{hc1} the parameter $\xi_2$ is constrained by \eqref{vincoli}). Therefore the CS-like terms must be absent both in $S_{bd}$ \eqref{Sbd} and in $S_{3D}$ \eqref{S3D}, in order to have $\TR$ symmetry preserved on the boundary.
\normalcolor

\section{Summary and discussion}\label{sec-summary}

In this paper we studied the consequences of the introduction of a flat boundary in the 4D covariant theory of fractons and LG \cite{Bertolini:2022ijb}, with the aim of investigating whether an induced 3D theory exists and, in that case, which is its physical meaning. Tightly related is the question of the existence of an algebraic structure on the boundary. The theory we are dealing with is not topological, and it is a common belief that only topological field theories show non-trivial boundary physics. Moreover, when a boundary is introduced in a QFT, the gauge symmetry plays a fundamental role, since it is the breaking of gauge invariance caused by the presence of the boundary that gives rise to an algebraic structure on the boundary which ``holographically'' induces a lower-dimensional gauge theory. The fracton symmetry \eqref{dA} is unusual, due to the presence of a double derivative, and considering a boundary on such a model $a\ priori$ has a non-trivial outcome. On the other hand it would not be the first case of a non-topological QFT exhibiting an induced theory on the boundary. In fact, this also happens in the case of Maxwell theory in 3D \cite{Maggiore:2019wie} and 4D \cite{Bertolini:2020hgr}, and we know that fracton models share many similarities with the electromagnetic theory \cite{Pretko:2016lgv,Bertolini:2022ijb}. Moreover, it has been shown that a fractonic $\theta$-term, which is a pure boundary term when $\theta$ is constant, gives rise to a 3D CS-like term and a  generalized Witten effect \cite{Pretko:2017xar}, with important consequences in condensed matter systems \cite{Pretko:2020cko,You:2019bvu}. 
A non-covariant CS-like term was studied in \cite{Cappelli:2015ocj}, where the higher-spin formalism is associated to dipolar behaviours in the context of Hall systems.
An algebraic structure on the boundary does exist indeed, as a consequence of the breaking of the Ward identities, and it can be interpreted  as a generalization of the standard $U(1)$ KM algebra, characterized by a double derivative, as it appears for instance also in \cite{You:2019bvu}. From the two broken Ward identities,
 the boundary DoF of the induced theory are identified as two symmetric traceless rank-2 tensors $\alpha_{ij}(X)$ and $\tilde\alpha_{ij}(X)$. It is worth to remark that
on the boundary some DoF disappear, since the boundary tensor fields turn out to be traceless. This might be due to the presence of a hidden symmetry, a guess that should be further investigated. 
The procedure to recover the induced theory leads to
the action $S_{3D}$ \eqref{S3D}, which is composed of a term similar to a higher-rank Maxwell contribution, written in terms of traceless rank-3 field strengths, which
mixes both fields $\alpha_{ij}(X)$ and $\tilde\alpha_{ij}(X)$, with a coefficient depending on the bulk constants $g_1$ and $g_2$, and a CS-like term for $\tilde\alpha_{ij}(X)$ with a free coefficient. 
Concerning the physical interpretation of our 3D induced theory $S_{3D}$ \eqref{S3D}, this can be identified with the ``traceless scalar charge'' model of fractons \cite{Pretko:2016lgv,Pretko:2016kxt,Du:2021pbc}. In fact the transformations of the boundary fields, the canonical commutators, the traceless conjugate momenta, $i.e.$ the ``electric fields'', coincide with what appears in the literature. 
This claim is confirmed also by the EoM of the 3D induced theory, from which two Gauss-like laws are derived, which imply the defining property of the fracton quasiparticles, $i.e.$ their limited mobility. Thus, one of the main results of this paper is that a non-standard covariant 3D traceless fracton theory turns out to be holographically induced from a 4D ordinary traceful covariant fracton theory. This claim gets even stronger confirmation from other components of the EoM, which can be identified with the Amp\`ere-like equations 
 of fractons \cite{Pretko:2017kvd}, further stressing the relation of fracton models with Maxwell theory. Concerning this analogy, we remark a close resemblance of our 3D action $S_{3D}$ \eqref{S3D} with Maxwell-Chern-Simons theory \cite{Deser:1981wh}, of which it appears to be a kind of spin-two generalization. A similar observation can also be found in \cite{Dalmazi3} in the context of self-dual massive gravity, where an identical covariant CS term appears, and whose relation with our 3D model is worth to be further investigated. However, differently from the standard Maxwell-Chern-Simons theory, in our paper all the coefficients are dimensionless, hence no topological mass can be identified. 
Therefore to better analyze this analogy, the study of the propagators would be helpful.
Notice also that the CS coefficient is free, thus it can be switched off. The choice of keeping the CS-like term or not is relevant for the physical interpretation of the model: by switching it off,  the 3D action $S_{3D}$ \eqref{S3D} can be decoupled into two Maxwell-like terms, and the boundary theory is compatible with $\TR$-symmetry, which characterizes the phenomenology involved. For instance, the physics on the boundary of the topological BF models \cite{Cho:2010rk,Cirio,Blasi:2011pf,Amoretti:2012hs} is identified with the effective description of the edge states of topological insulators, where $\TR$ is preserved both on the bulk and on the boundary. On the other hand, keeping the CS-like term, $i.e.$ relaxing the $\TR$ constraint, the EoM get a matter contribution. In particular the CS-like term plays the role of fractonic charge $\tilde\rho(X)$ \eqref{rho} and current $\tilde J_{\textsc{ij}}(X)$ \eqref{J} in two of the Maxwell-like equations, in accordance with \cite{Pretko:2017xar}.
Some final physical remarks are in order. Differently from the standard electromagnetic theory and the 4D traceful fracton model, here the magnetic-like vectors $B_\textsc{a}(X)$ and $\tilde B_\textsc{a}(X)$ do not have zero divergence, nor a Bianchi identity exists for the traceless rank-3 field strengths $\varphi_{abc}(X),\ \tilde\varphi_{abc}(X)$, which suggests the presence of fractonic 3D vortices. Additionally, 3D fracton models are known to be related to the elasticity theory of topological defects through a duality \cite{Pretko:2017kvd}. For instance the traceful Amp\`ere-like equation can be seen as describing the motion of these defects. Under this respect, it would be interesting to understand if and how our traceless boundary theory can be related to topological defects. Finally, there seems to be an interesting possible interpretation of the fractonic CS-like term as associated to torsion contributions, as in \cite{Gromov:2017vir}, which also would be worth to further analyze.

\section*{Acknowledgments}

We thank Alberto Blasi and Andrea Damonte for enlightening discussions. This work has been partially supported by the INFN Scientific Initiative GSS: ``Gauge Theory, Strings and Supergravity''. E.B. is supported by MIUR grant ``Dipartimenti di Eccellenza'' (100020-2018-SD-DIP-ECC\_001).

\appendix

\section{Commutators}

\subsection{The bulk: generalized Ka\c{c}-Moody algebra}\label{appAlgebra}

	Considering the first Ward identity \eqref{wi1} 
	\be\label{wi1A}
	\int dx^3\theta(x^3)\partial_i\partial_jJ^{ij}=2(g_2-g_1)\partial_i\partial_j\tilde A^{ij}-2g_2\partial_i\partial^i\tilde A|_{x^3=0}\ ,
	\ee
	we compute \\
	$\frac{\delta}{\delta J^{mn}(x')}\eqref{wi1A}$ :
	\be
	\begin{split}
	\partial_m\partial_n\delta^{(3)}(X-X')&=2(g_2-g_1)\partial_i\partial_j\frac{\delta^2Z_c[J,\tilde J]}{\delta J^{mn}(X')\tilde J_{ij}(X)}-2g_2\eta_{ij}\partial_a\partial^a\frac{\delta^2Z_c[J,\tilde J]}{\delta J^{mn}(X')\tilde J_{ij}(X)}\\
	&=2\left[(g_2-g_1)\delta^k_i\delta^l_j-g_2\eta^{kl}\eta_{ij}\right]\partial_k\partial_l\frac{\delta^2Z_c[J,\tilde J]}{\delta J^{mn}(X')\tilde J_{ij}(X)}\\
	&=2i\left[(g_2-g_1)\delta^k_i\delta^l_j-g_2\eta^{kl}\eta_{ij}\right]\partial_k\partial_l\langle T(A_{mn}(X')\tilde A^{ij}(X))\rangle\\
	&=2i\left[(g_2-g_1)\delta^k_i\delta^l_j-g_2\eta^{kl}\eta_{ij}\right]\cancel{\langle T(A_{mn}(X')\partial_k\partial_l\tilde A^{ij})\rangle}+\\
&\quad+2i\left[(g_2-g_1)\delta^k_i\delta^l_j-g_2\eta^{kl}\eta_{ij}\right]\left\{\left[\partial_l\tilde A^{ij}(X),A_{mn}(X')\right]\delta^0_k\delta(x^0-x'^0)+\right.\\
&\quad+\left.\partial_k\left(\left[\tilde A^{ij}(X),A_{mn}(X')\right]\delta^0_l\delta(x^0-x'^0)\right)\right\}\\
	&=2i\left[(g_2-g_1)\left(\partial_j\tilde A^{0j}+\partial_\textsc{a}\tilde A^{0\textsc{a}}\right)-g_2\partial^0\tilde A\ ,\ A'_{mn}\right]\delta(x^0-x'^0)+\\
	&\quad+2i\partial_0\left\{\left[(g_2-g_1)\tilde A^{00}+g_2\tilde A\ ,\ A'_{mn}\right]\delta(x^0-x'^0)\right\}\ ,
	\end{split}
	\ee
where we used the conserved current equation \eqref{cc1}. Integrating over $dx^0$, we finally get to the following equal time commutators
	\begin{empheq}{align}
	\left[\Delta\tilde A(X)\ ,\ A_{0n}(X')\right]_{x^0=x'^0}&=0\label{app[DAt,A0n]}\\
	\left[\Delta\tilde A(X)\ ,\ A_{\textsc{mn}}(X')\right]_{x^0=x'^0}&=i\partial_\textsc{m}\partial_\textsc{n}\{\delta(x^1-x'^1)\delta(x^2-x'^2)\}\ ,\label{comm1}
	\end{empheq}
where we defined
	\be\label{DA't}
	\Delta \tilde A\equiv2(g_1-g_2)\left(\partial_j\tilde A^{0j}+\partial_\textsc{a}\tilde A^{0\textsc{a}}\right)+2g_2\partial^0\tilde A\ .
	\ee
In the same way, we now compute
	$\frac{\delta}{\delta \tilde J^{mn}(x')}\eqref{wi1A}$ :
	\be
	\begin{split}
	0&=2(g_2-g_1)\partial_i\partial_j\frac{\delta^2Z_c[J,\tilde J]}{\delta\tilde J^{mn}(X')\tilde J_{ij}(X)}-2g_2\eta_{ij}\partial_a\partial^a\frac{\delta^2Z_c[J,\tilde J]}{\delta\tilde J^{mn}(X')\tilde J_{ij}(X)}\\
	&=2i\left[(g_2-g_1)\left(\partial_j\tilde A^{0j}+\partial_\textsc{a}\tilde A^{0\textsc{a}}\right)-g_2\partial^0\tilde A\ ,\ \tilde A'_{mn}\right]\delta(x^0-x'^0)+\\
	&\quad+2i\partial_0\left\{\left[(g_2-g_1)\tilde A^{00}+g_2\tilde A\ ,\ \tilde A'_{mn}\right]\delta(x^0-x'^0)\right\}\ ,
	\end{split}
	\ee
where we used again the conserved current equation \eqref{cc1}. By integrating over time and using the definition \eqref{DA't}, we find the following equal time commutator
	\be
	\left[\Delta\tilde A(X)\ ,\ \tilde A_{mn}(X')\right]_{x^0=x'^0}=0\ .
	\ee
Taking the second broken Ward identity \eqref{wi2}
	\be\label{wi2A}
	\partial_i\partial_j\tilde J^{ij}|_{x^3=0}=-2(g_2-g_1)\partial_i\partial_j A^{ij}+2g_2\partial_i\partial^iA|_{x^3=0}\ .
	\ee
we compute 	$\frac{\delta}{\delta J^{mn}(x')}\eqref{wi2A}$ :
	\be
	\begin{split}
	0&=-2(g_2-g_1)\partial_i\partial_j\frac{\delta^2Z_c[J,\tilde J]}{\delta J^{mn}(X') J_{ij}(X)}+2g_2\eta_{ij}\partial_a\partial^a\frac{\delta^2Z_c[J,\tilde J]}{\delta J^{mn}(X') J_{ij}(X)}\\
	&=-2i\left[(g_2-g_1)\left(\partial_j A^{0j}+\partial_\textsc{a} A^{0\textsc{a}}\right)-g_2\partial^0 A\ ,\  A'_{mn}\right]\delta(x^0-x'^0)+\\
	&\quad-2i\partial_0\left\{\left[(g_2-g_1)  A^{00}+g_2 A\ ,\  A'_{mn}\right]\delta(x^0-x'^0)\right\}\ ,
	\end{split}
	\ee
where we used \eqref{cc2}. Integrating over $dx^0$ we find the equal time commutator
	\be
	\left[\Delta A(X)\ ,\  A_{mn}(X')\right]_{x^0=x'^0}=0\ ,
	\ee
where $\Delta A(X)$ is defined as  \eqref{DA't}
	\be\label{DA'}
	\Delta A\equiv2(g_1-g_2)\left(\partial_jA^{0j}+\partial_\textsc{a}A^{0\textsc{a}}\right)+2g_2\partial^0A\ .
	\ee
We finally compute $\frac{\delta}{\delta\tilde J^{mn}(x')}\eqref{wi2A}$ :
	\be
	\begin{split}
	\partial_m\partial_n\delta^{(3)}(X-X')&=-2(g_2-g_1)\partial_i\partial_j\frac{\delta^2Z_c[J,\tilde J]}{\delta\tilde J^{mn}(X') J_{ij}(X)}+2g_2\eta_{ij}\partial_a\partial^a\frac{\delta^2Z_c[J,\tilde J]}{\delta\tilde J^{mn}(X') J_{ij}(X)}\\
	&=-2i\left[(g_2-g_1)\left(\partial_j A^{0j}+\partial_\textsc{a} A^{0\textsc{a}}\right)-g_2\partial^0 A\ ,\ \tilde A'_{mn}\right]\delta(x^0-x'^0)+\\
	&\quad-2i\partial_0\left\{\left[(g_2-g_1) A^{00}+g_2 A\ ,\ \tilde A'_{mn}\right]\delta(x^0-x'^0)\right\}\ ,
	\end{split}
	\ee
where we used \eqref{cc2} and from which, integrating over $dx^0$, we find
	\begin{empheq}{align}
	\left[\Delta A(X)\ ,\ \tilde A_{0n}(X')\right]_{x^0=x'^0}&=0\\
    	\left[-\Delta A(X)\ ,\ \tilde A_{\textsc{mn}}(X')\right]_{x^0=x'^0}&=i\partial_\textsc{m}\partial_\textsc{n}\{\delta(x^1-x'^1)\delta(x^2-x'^2)\}\ .\label{comm2}
	\end{empheq}
	
\subsection{The boundary: canonical commutators}\label{appComm}

We take the commutator \eqref{[DAt,A]}  and its trace, in the following equal time combination
	\be\label{DAt,A-TrA appB}
	\left[\Delta\tilde A\ ,\ A_{\textsc{df}}'-\frac{1}{2}\eta_{\textsc{df}}\eta^{\textsc{mn}}A_{\textsc{mn}}'\right]=\frac{i}{2}\left(\delta^{\textsc{m}}_{\textsc{d}}\delta^{\textsc{n}}_{\textsc{f}}+\delta^{\textsc{n}}_{\textsc{d}}\delta^{\textsc{m}}_{\textsc{f}}-\eta^{\textsc{mn}}\eta_{\textsc{df}}\right)\partial_\textsc{m}\partial_\textsc{n}\delta^{(2)}(X-X')\ .
	\ee
In terms of the solutions on the boundary \eqref{a,at sol} we have
	\begin{empheq}{align}
	 A_{\textsc{df}}(X)|_\eqref{sol2}&=\epsilon_{\textsc{d}ab}\partial^a \alpha^b_{\ \textsc{f}}(X)+\epsilon_{\textsc{f}ab}\partial^a \alpha^b_{\ \textsc{d}}(X)\label{sol2'}\\
	 \eta^{\textsc{mn}}A_{\textsc{mn}}|_\eqref{sol2}&=2\epsilon_{0\textsc{bc}}\partial^\textsc{b}\alpha^{0\textsc{c}}=A_{00}(X)|_\eqref{sol2}\label{TrA=A00}\\
	\Delta \tilde A|_\eqref{sol1}&=\partial_\textsc{m}\partial_\textsc{n}\left[g_{12}\left(\epsilon^{0\textsc{mb}}\tilde \alpha_{\textsc{b}}^{\ \textsc{n}}+\epsilon^{0\textsc{nb}}\tilde \alpha_{\textsc{b}}^{\ \textsc{m}}\right)\right]\ ,\label{DAt-sol}
	\end{empheq}
where $g_{12}\equiv2(g_1-g_2)$. As a consequence of the tracelessness of $A_{ij}(X)$, we can use \eqref{TrA=A00} and write the commutator \eqref{app[DAt,A0n]} for $n=0$ as follows
	\be\label{DA,TrA=0}
	\left[\Delta\tilde A(X)|_\eqref{sol1}\ ,\ A_{00}(X')|_\eqref{sol2}\right]=\left[\Delta\tilde A(X)|_\eqref{sol1}\ ,\ \eta^{\textsc{mn}}A_{\textsc{mn}}(X')|_\eqref{sol2}\right]=0\ ,
	\ee
then, using \eqref{sol2'},  \eqref{DAt-sol} and \eqref{DA,TrA=0}, the commutator \eqref{DAt,A-TrA appB} becomes
	\be
	\begin{split}
	\frac{i}{2}\partial_\textsc{m}\partial_\textsc{n}\left\{...
	\right\}^{\textsc{mn}}_{\textsc{df}}&=\left[\Delta\tilde A(X)|_\eqref{sol1}\ ,\ A_{\textsc{df}}(X')|_\eqref{sol2}\right]\\
	&=\partial_\textsc{m}\partial_\textsc{n}\left[g_{12}\left(\epsilon^{0\textsc{mb}}\tilde \alpha_{\textsc{b}}^{\ \textsc{n}}(X)+\epsilon^{0\textsc{nb}}\tilde \alpha_{\textsc{b}}^{\ \textsc{m}}(X)\right)\ ,\ \epsilon_{\textsc{d}ab}\partial^a \alpha^b_{\ \textsc{f}}(X')+\epsilon_{\textsc{f}ab}\partial^a \alpha^b_{\ \textsc{d}}(X')\right]\ ,
	\end{split}
	\ee
from which we can identify the following canonical commutation relation
	
	\be\label{P,Q}
	\left[Q^{\textsc{mn}}\ , \ P'_{\textsc{df}}\right]=i\left(\frac{\delta^{\textsc{m}}_{\textsc{d}}\delta^{\textsc{n}}_{\textsc{f}}+\delta^{\textsc{n}}_{\textsc{d}}\delta^{\textsc{m}}_{\textsc{f}}}{2}-\frac{1}{2}\eta^{\textsc{mn}}\eta_{\textsc{df}}\right)\delta^{(2)}(X-X')\ ,
	\ee
with
	\begin{empheq}{align}
	Q^{\textsc{mn}}=Q^{\textsc{nm}}&\equiv\epsilon^{0\textsc{mb}}\tilde \alpha_{\textsc{b}}^{\ \textsc{n}}+\epsilon^{0\textsc{nb}}\tilde \alpha_{\textsc{b}}^{\ \textsc{m}} \\
	P_{\textsc{df}}=P_{\textsc{fd}}\ &\equiv g_{12}\left(\epsilon_{\textsc{d}ab}\partial^a \alpha^b_{\ \textsc{f}}+\epsilon_{\textsc{f}ab}\partial^a \alpha^b_{\ \textsc{d}}\right)
	\end{empheq}
and $Q^\textsc{m}_{\ \textsc{m}}=P^\textsc{m}_{\ \textsc{m}}=0$. We can go further, multiplying both right and left hand sides of \eqref{P,Q} by \mbox{$\epsilon_{0\textsc{ma}}\epsilon^{0\textsc{fk}}=\delta_\textsc{m}^\textsc{k}\delta_\textsc{a}^\textsc{f}-\delta_\textsc{m}^\textsc{f}\delta_\textsc{a}^\textsc{k}$}
	\be
	\left[\epsilon_{0\textsc{ma}}Q^{\textsc{mn}}\ , \ \epsilon^{0\textsc{fk}}P'_{\textsc{df}}\right]=i(\delta_\textsc{m}^\textsc{k}\delta_\textsc{a}^\textsc{f}-\delta_\textsc{m}^\textsc{f}\delta_\textsc{a}^\textsc{k})\left(\frac{\delta^{\textsc{m}}_{\textsc{d}}\delta^{\textsc{n}}_{\textsc{f}}+\delta^{\textsc{n}}_{\textsc{d}}\delta^{\textsc{m}}_{\textsc{f}}}{2}-\frac{1}{2}\eta^{\textsc{mn}}\eta_{\textsc{df}}\right)\delta^{(2)}(X-X')\ ,
	\ee
for which
	\be
	\begin{split}
	\epsilon_{0\textsc{ma}}Q^{\textsc{mn}}&=\epsilon_{0\textsc{ma}}\left(\epsilon^{0\textsc{mb}}\tilde \alpha_{\textsc{b}}^{\ \textsc{n}}+\epsilon^{0\textsc{nb}}\tilde \alpha_{\textsc{b}}^{\ \textsc{m}}\right)\\
	&=-2\tilde \alpha_\textsc{a}^{\ \textsc{n}}+\delta^\textsc{n}_\textsc{a}\tilde \alpha^\textsc{m}_{\ \textsc{m}}
	\end{split}
	\ee
which is the traceless spatial part of $\tilde \alpha_{{mn}}(X) $. Then
	\be
	\begin{split}
	\epsilon^{0\textsc{fk}}P_{\textsc{df}}&=g_{12}\left[(\delta^0_a\delta^\textsc{k}_b-\delta^0_b\delta^\textsc{k}_a)\partial^a\alpha^b_{\ \textsc{d}}+\epsilon^{0\textsc{fk}}\epsilon_{0\textsc{da}}\partial^\textsc{a}\alpha^0_{\ \textsc{f}}-\epsilon^{0\textsc{fk}}\epsilon_{0\textsc{db}}\partial^0\alpha^\textsc{b}_{\ \textsc{f}}\right]\\
	&=g_{12}\left[ 2(\partial^0\alpha^\textsc{k}_{\ \textsc{d}}-\partial^\textsc{k}\alpha^0_{\ \textsc{d}})+\delta^\textsc{k}_\textsc{d}(\partial^\textsc{a}\alpha^0_{\ \textsc{a}}-\partial^0\alpha^\textsc{b}_{\ \textsc{b}})\right]\ .
	\end{split}
	\ee
Finally, at the right hand side we have
	\be
	(\delta_\textsc{m}^\textsc{k}\delta_\textsc{a}^\textsc{f}-\delta_\textsc{m}^\textsc{f}\delta_\textsc{a}^\textsc{k})\left(\frac{\delta^{\textsc{m}}_{\textsc{d}}\delta^{\textsc{n}}_{\textsc{f}}+\delta^{\textsc{n}}_{\textsc{d}}\delta^{\textsc{m}}_{\textsc{f}}}{2}-\frac{1}{2}\eta^{\textsc{mn}}\eta_{\textsc{df}}\right)=\frac{1}{2}\left(\delta^\textsc{n}_\textsc{a}\delta^\textsc{k}_\textsc{d}-\delta^\textsc{n}_\textsc{d}\delta^\textsc{k}_\textsc{a}-\eta_{\textsc{ad}}\eta^{\textsc{kn}}\right)\ .
	\ee
By properly raising and lowering the indices with {$\eta^{\textsc{dl}}\eta_{\textsc{nb}}$}, we finally get
	\be
	g_{12}\biggl[2\tilde \alpha_\textsc{ab}-\eta_\textsc{ab}\tilde \alpha^\textsc{m}_{\ \textsc{m}}\ ,\  2(\partial^0\alpha^\textsc{kl}-\partial^\textsc{k}\alpha^{0\textsc{l}})'+\eta^\textsc{kl}(\partial_\textsc{f}\alpha^{0\textsc{f}}-\partial^0\alpha^\textsc{f}_{\ \textsc{f}})'\biggr]=\frac{i}{2}\left(\delta^{\textsc{k}}_{\textsc{a}}\delta^{\textsc{l}}_{\textsc{b}}+\delta^{\textsc{l}}_{\textsc{a}}\delta^{\textsc{k}}_{\textsc{b}}-\eta^{\textsc{kl}}\eta_{\textsc{ab}}\right)\delta^{(2)}(X-X')\ ,
	\ee
where the primed quantities  depend on $X'$. At the right hand side we have the index symmetry $\textsc{a}\leftrightarrow \textsc{b}$ and $\textsc{c}\leftrightarrow \textsc{d}$, while at the left hand side the symmetry is only for  $\textsc{a}\leftrightarrow \textsc{b}$. We thus symmetrize the result as follows
	\be
	\frac{1}{2}\left(\biggl[..._{\textsc{ab}}\ ,\ ...^\textsc{cd}\biggr]+\biggl[..._{\textsc{ab}}\ ,\ ...^\textsc{dc}\biggr]\right)=\frac{i}{2}\left(\delta^{\textsc{c}}_{\textsc{a}}\delta^{\textsc{d}}_{\textsc{b}}+\delta^{\textsc{d}}_{\textsc{a}}\delta^{\textsc{c}}_{\textsc{b}}-\eta^{\textsc{cd}}\eta_{\textsc{ab}}\right)\delta^{(2)}(X-X')\ ,
	\ee
obtaining
	\be
	\left[\tilde \alpha _{\textsc{ab}}-\frac{1}{2}\eta_{\textsc{ab}}\tilde \alpha ^\textsc{m}_{\ \textsc{m}}\ ,\ -g_{12}\left(2f'^{\textsc{cd}0}-\eta^{\textsc{cd}}f'^{{\ a}0}_{a}\right)\right]=\frac{i}{2}\left(\delta^{\textsc{c}}_{\textsc{a}}\delta^{\textsc{d}}_{\textsc{b}}+\delta^{\textsc{d}}_{\textsc{a}}\delta^{\textsc{c}}_{\textsc{b}}-\eta^{\textsc{cd}}\eta_{\textsc{ab}}\right)\delta^{(2)}(X-X')\ ,\label{qt,p}
	\ee
where $f_{abc}(X)$ is analogous to $F_{\mu\nu\rho}(x)$ \eqref{Fmunurho}, but referred to $\alpha_{ab}(X)$, $i.e.$
	\be\label{fabc}
	f_{abc}\equiv\partial_a\alpha_{bc}+\partial_b\alpha_{ac}-2\partial_c\alpha_{ab}\ .
	\ee
From \eqref{qt,p} we can identify the new canonical variables as
	\begin{empheq}{align}
	\tilde q_{\textsc{ab}}&\equiv\tilde \alpha _{\textsc{ab}}-\frac{1}{2}\eta_{\textsc{ab}}\tilde \alpha ^\textsc{m}_{\ \textsc{m}}\\
	\tilde p^{\textsc{cd}}&\equiv -2g_{12}\left(f^{\textsc{cd}0}-\frac{1}{2}\eta^{\textsc{cd}}f^{\ a0}_{a}\right)=-2g_{12}\left[(\partial^\textsc{c}\alpha^{0\textsc{d}}+\partial^\textsc{d}\alpha^{0\textsc{c}}-2\partial^0\alpha^{\textsc{cd}})-\eta^{\textsc{cd}}\partial_a\alpha^{0a}\right]\ .
	\end{empheq} 
Not surprisingly, starting from the commutator \eqref{[DA,At]}, and proceeding as we just did, we land on an analogous result (up to a sign) with $\alpha\leftrightarrow\tilde \alpha$ switched, $i.e.$ we get
	\be
	\left[\alpha _{\textsc{ab}}-\frac{1}{2}\eta_{\textsc{ab}}\alpha ^\textsc{m}_{\ \textsc{m}}\ ,\ g_{12}\left(2\tilde f'^{\textsc{cd}0}-\eta^{\textsc{cd}}\tilde f'^{{\ a}0}_{a}\right)\right]=\frac{i}{2}\left(\delta^{\textsc{c}}_{\textsc{a}}\delta^{\textsc{d}}_{\textsc{b}}+\delta^{\textsc{d}}_{\textsc{a}}\delta^{\textsc{c}}_{\textsc{b}}-\eta^{\textsc{cd}}\eta_{\textsc{ab}}\right)\delta^{(2)}(X-X')\ ,\label{q,pt}
	\ee
where $\tilde f_{abc}(X)$ refers to $\tilde\alpha_{ab}(X)$
	\be
	\tilde f_{abc}\equiv\partial_a\tilde \alpha_{bc}+\partial_b\tilde \alpha_{ac}-2\partial_c\tilde \alpha_{ab}\ .
	\ee
From \eqref{q,pt} we can identify another set of canonical variables
	\begin{empheq}{align}
	q_{\textsc{ab}}&\equiv \alpha _{\textsc{ab}}-\frac{1}{2}\eta_{\textsc{ab}} \alpha ^\textsc{m}_{\ \textsc{m}}\\
	p^{\textsc{cd}}&\equiv 2g_{12}\left(\tilde f^{\textsc{cd}0}-\frac{1}{2}\eta^{\textsc{cd}}\tilde f^{\ a0}_{a}\right)=2g_{12}\left[(\partial^\textsc{c}\tilde\alpha^{0\textsc{d}}+\partial^\textsc{d}\tilde\alpha^{0\textsc{c}}-2\partial^0\tilde\alpha^{\textsc{cd}})-\eta^{\textsc{cd}}\partial_a\tilde\alpha^{0a}\right]\ .
	\end{empheq}
We observe that the canonical variables $\tilde q_{\textsc{ab}}(X),\ q_{\textsc{ab}}(X)$ in \eqref{qt,p} and \eqref{q,pt}, depend on the traceless spatial part of the fields $\tilde\alpha_{ij}(X)$ and $\alpha_{ij}(X)$.

\section{The most general action}\label{appSgen}

The most general action of the 3D boundary theory must be compatible with
	\bi
	\item power-counting $[\alpha]=0,\ [\tilde\alpha]=1$ ;
	\item symmetry $\delta S=\tilde\delta S=0$, where $\delta$ and $\tilde\delta$ are defined in \eqref{dalpha} and \eqref{dalphat} ;
	\item canonical variables, $i.e.\ \frac{\partial\mathcal L_{kin}}{\partial \dot q}=p$, identified in \eqref{al,ft} and \eqref{alt,f}.
	\ei
From the first two requests we have that the most general action must be
	\be \label{appS3D-gen}
		\begin{split}
		S_{3D}&=\int d^3X\left\{\omega_1\left(\partial_c\alpha_{ab}\partial^c\alpha^{ab}-\frac{3}{2}\partial_c\alpha_{ab}\partial^a\alpha^{bc}\right)+\omega_2\left(\partial_c\alpha_{ab}\partial^c\tilde\alpha^{ab}-\frac{3}{2}\partial_c\alpha_{ab}\partial^a\tilde\alpha^{bc}\right)-\right.\\
		&\quad\left.-3\omega_3\,\alpha^d_a\epsilon^{abc}\partial_b\alpha_{cd}-3\omega_4\,\tilde\alpha^d_a\epsilon^{abc}\partial_b\alpha_{cd}-3\omega_5\,\tilde\alpha^d_a\epsilon^{abc}\partial_b\tilde\alpha_{cd}\right\}\\
		&=\int d^3X\left\{\frac{\omega_1}{6}\,\varphi_{abc}\varphi^{abc}+\frac{\omega_2}{6}\,\varphi_{abc}\tilde\varphi^{abc}+\omega_3\, \alpha^d_a\epsilon^{abc}\varphi_{dbc}+\omega_4\, \alpha^d_a\epsilon^{abc}\tilde\varphi_{dbc}+\omega_5\, \tilde\alpha^d_a\epsilon^{abc}\tilde\varphi_{dbc}\right\}
		\end{split}
	\ee
where $\omega_i$ are constants, $[\omega_2]=[\omega_5]=0,\ [\omega_1]=[\omega_4]=1,\ [\omega_3]=2$, and we defined the tensor
	\be\label{appphi}
		\begin{split}
		\varphi_{abc}&\equiv f_{abc}+\frac{1}{4}\left(-2\eta_{ab}f^d_{\ dc}+\eta_{bc}f^d_{\ da}+\eta_{ac}f^d_{\ db}\right)\\
		&=-2\partial_c\alpha_{ab}+\partial_a\alpha_{bc}+\partial_b\alpha_{ac}-\eta_{ab}\partial^d\alpha_{dc}+\frac{1}{2}\eta_{bc}\partial^d\alpha_{da}+\frac{1}{2}\eta_{ac}\partial^d\alpha_{db}\ ,
		\end{split}
	\ee
and its analog $\tilde\varphi_{abc}(X)$ with respect to $\tilde\alpha_{ab}(X)$, with the following properties
	\begin{empheq}{align}
	&\varphi_{abc}=\varphi_{bac}\quad;\quad	\tilde\varphi_{abc}=\tilde\varphi_{bac}\\
	&\varphi_{abc}+\varphi_{cab}+\varphi_{bca}=0=\tilde\varphi_{abc}+\tilde\varphi_{cab}+\tilde\varphi_{bca}\label{appcicl}\\
	&\delta\varphi_{abc}=\tilde\delta\tilde\varphi_{abc}=0\\
	&\eta^{ab}\varphi_{abc}=\eta^{bc}\varphi_{abc}=\eta^{ab}\tilde\varphi_{abc}=\eta^{bc}\tilde\varphi_{abc}=0\ .\label{apptraceless}
	\end{empheq}
Notice that 
	 \begin{empheq}{align}
 	\tilde\varphi^{\textsc{mn}0}&=\tilde f^{\textsc{mn}0}-\frac{1}{2}\eta^{\textsc{mn}}\tilde f_a^{\ a0}=\frac{1}{2g_{12}}p^{\textsc{mn}}\label{appp}\\
 	\varphi^{\textsc{mn}0}&=f^{\textsc{mn}0}-\frac{1}{2}\eta^{\textsc{mn}} f_a^{\ a0}=-\frac{1}{2g_{12}}\tilde p^{\textsc{mn}}\ .
	\end{empheq}
We rewrite the fields $\alpha_{ab}(X)$ and $\tilde\alpha_{ab}(X)$ according to the representation of the rotation group, as follows
	\begin{empheq}{align}
	\alpha_{00}&=-4\psi=\alpha^\textsc{a}_\textsc{a}\label{apppsi}\\
	\alpha_{0\textsc{a}}&=v_{\textsc{a}}\label{appv}\\
	\alpha_{\textsc{ab}}&=2s_{\textsc{ab}}-2\eta_{\textsc{ab}}\psi\quad;\quad s_{\textsc{ab}}\equiv\frac{1}{2}\left(\alpha_{\textsc{ab}}-\frac{1}{2}\eta_{\textsc{ab}}\alpha^\textsc{d}_\textsc{d}\right)=\frac{1}{2}q_{\textsc{ab}}\label{apps}
	\end{empheq}
and
	\begin{empheq}{align}
	\tilde\alpha_{00}&=-4\tilde\psi=\tilde\alpha^\textsc{a}_\textsc{a}\\
	\tilde\alpha_{0\textsc{a}}&=\tilde v_{\textsc{a}}\\
	\tilde\alpha_{\textsc{ab}}&=2\tilde s_{\textsc{ab}}-2\eta_{\textsc{ab}}\tilde\psi\quad;\quad \tilde s_{\textsc{ab}}\equiv\frac{1}{2}\left(\tilde\alpha_{\textsc{ab}}-\frac{1}{2}\eta_{\textsc{ab}}\tilde\alpha^\textsc{d}_\textsc{d}\right)=\frac{1}{2}\tilde q_{\textsc{ab}}\ ,
	\end{empheq}
in terms of which the commutators \eqref{al,ft} and \eqref{alt,f} become
	\begin{empheq}{align}
	\left[2s_{\textsc{ab}}\ ,\ 2g_{12}\tilde\varphi'^{\textsc{cd}0}\right]=&\frac{i}{2}\left(\delta^{\textsc{c}}_{\textsc{a}}\delta^{\textsc{d}}_{\textsc{b}}+\delta^{\textsc{d}}_{\textsc{a}}\delta^{\textsc{c}}_{\textsc{b}}-\eta_{\textsc{ab}}\eta^{\textsc{cd}}\right)\delta^{(2)}(X-X')\label{appq,p}\\
	\left[2\tilde s_{\textsc{ab}}\ ,\ -2g_{12}\varphi'^{\textsc{cd}0}\right]=&\frac{i}{2}\left(\delta^{\textsc{c}}_{\textsc{a}}\delta^{\textsc{d}}_{\textsc{b}}+\delta^{\textsc{d}}_{\textsc{a}}\delta^{\textsc{c}}_{\textsc{b}}-\eta_{\textsc{ab}}\eta^{\textsc{cd}}\right)\delta^{(2)}(X-X')\ .\label{appqt,pt}
	\end{empheq}
Compatibility of the action $S_{3D}$ \eqref{appS3D-gen} with the first commutator \eqref{appq,p}, $i.e.$
	\be\label{appLkin}
	\frac{\partial\mathcal L_{3D}}{\partial \dot q_{\textsc{ab}}}=p^{\textsc{ab}}\ ,
	\ee
requires
	\be
	\omega_1=\omega_3=\omega_4=0\ ,
	\ee
and $\omega_5$ is left free. Finally, distinguishing between time and space indices in the term of \eqref{appS3D-gen} which has $\omega_2$ as coefficient, we have
	\be
	\varphi_{abc}\tilde\varphi^{abc}=\frac{3}{2}\varphi_{00\textsc{a}}\tilde\varphi^{00\textsc{a}}+2\varphi_{0\textsc{ab}}\tilde\varphi^{0\textsc{ab}}+\varphi_{\textsc{ab}0}\tilde\varphi^{\textsc{ab}0}+\varphi_{\textsc{abc}}\tilde\varphi^{\textsc{abc}}\ ,
	\ee
where we observed that $\varphi_{\textsc{a}00}=-\frac{1}{2}\varphi_{00\textsc{a}}$ (and the same for $\tilde\varphi_{\textsc{a}00}$). Additionally, by using \eqref{apppsi}, \eqref{appv} and \eqref{apps}, we have
	\begin{empheq}{align}
	\varphi_{00\textsc{a}}&=6\partial_\textsc{a}\psi+\partial_0v_\textsc{a}+\partial^\textsc{b}q_{\textsc{ab}}\\
	\varphi_{\textsc{ab}0}&=-2\partial_0q_{\textsc{ab}}+\partial_\textsc{a}v_\textsc{b}+\partial_\textsc{b}v_\textsc{a}-\eta_{\textsc{ab}}\partial^\textsc{d}v_\textsc{d}\\
	\varphi_{0\textsc{ab}}&=-2\partial_\textsc{b}v_\textsc{a}+\partial_\textsc{a}v_\textsc{b}+\partial_0q_{\textsc{ab}}+\tfrac{1}{2}\eta_{\textsc{ab}}\partial^\textsc{d}v_\textsc{d}\\
	\varphi_{\textsc{abc}}&=-2\partial_\textsc{c}q_{\textsc{ab}}+\partial_\textsc{b}q_{\textsc{ac}}+\partial_\textsc{a}q_{\textsc{bc}}+6\eta_{\textsc{ab}}\partial_\textsc{c}\psi-3\eta_{\textsc{bc}}\partial_\textsc{a}\psi-3\eta_{\textsc{ac}}\partial_\textsc{b}\psi-\\
	&\quad-\partial^0(\eta_{\textsc{ab}}v_\textsc{c}-\tfrac{1}{2}\eta_{\textsc{ac}}v_\textsc{b}-\tfrac{1}{2}\eta_{\textsc{bc}}v_\textsc{a})-\partial^\textsc{d}(\eta_{\textsc{ab}}q_{\textsc{dc}}-\tfrac{1}{2}\eta_{\textsc{ac}}q_{\textsc{db}}-\tfrac{1}{2}\eta_{\textsc{bc}}q_{\textsc{da}})\ .\nonumber
	\end{empheq}
The only terms containing $\dot q$ contributions are related to $\varphi_{\textsc{ab}0}$ and $\varphi_{0\textsc{ab}}$, $i.e.$ in the 3D Lagrangian they only appear in $\frac{\omega_2}{6}(2\varphi_{0\textsc{ab}}\tilde\varphi^{0\textsc{ab}}+\varphi_{\textsc{ab}0}\tilde\varphi^{\textsc{ab}0})$. Keeping in mind this, consistently with its definition, for a traceless tensor in $d$ dimensions we have
	\be\label{apptraceless-der}
	\frac{\partial s_{\mu\nu}}{\partial s_{\alpha\beta}}=\frac{\delta^\alpha_\mu\delta^\beta_\nu+\delta^\alpha_\nu\delta^\beta_\mu}{2}-\frac{1}{d}\eta^{\alpha\beta}\eta_{\mu\nu}\ ,
	\ee
the compatibility condition \eqref{appLkin} implies
	\be
		\begin{split}
		2g_{12}\tilde\varphi^{\textsc{ab}0}&=\frac{\partial\mathcal L_{3D}}{\partial \dot q_{\textsc{ab}}}\\
		&=\frac{\omega_2}{6}\left(2\frac{\partial\varphi_{0\textsc{mn}}}{\partial \dot q_{\textsc{ab}}}\tilde\varphi^{0\textsc{mn}}+\frac{\partial\varphi_{\textsc{mn}0}}{\partial \dot q_{\textsc{ab}}}\tilde\varphi^{\textsc{mn}0}\right)\\
		&=\frac{\omega_2}{6}\left(-2\tilde\varphi^{\textsc{ab}0}+\tilde\varphi^{0\textsc{ab}}+\tilde\varphi^{0\textsc{ba}}\right)\\
		&=-\frac{\omega_2}{2}\tilde\varphi^{\textsc{ab}0}\ ,
		\end{split}
	\ee
due to the cyclicity of $\tilde\varphi_{abc}(X)$ \eqref{appcicl}. We thus find
	\be
	\omega_2=-4g_{12}\ ,
	\ee
from which the 3D action \eqref{appS3D-gen} becomes
	\be \label{appS3D}
		\begin{split}
		S_{3D}&=\int d^3X\left[-4g_{12}\left(\partial_c\alpha_{ab}\partial^c\tilde\alpha^{ab}-\frac{3}{2}\partial_c\alpha_{ab}\partial^a\tilde\alpha^{bc}\right)+3\omega_5\,\tilde\alpha^d_a\epsilon^{abc}\partial_b\tilde\alpha_{cd}\right]\\
		&=\int d^3X\left[-g_{12}\left(\frac{2}{3}f_{abc}\tilde f^{abc}-\frac{1}{2}f_a^{\ ab}\tilde f^c_{\ cb}\right)+\omega_5\, \tilde\alpha^d_a\epsilon^{abc}\tilde f_{dbc}\right]\\
		&=\int d^3X\left(-\frac{2}{3}g_{12}\,\varphi_{abc}\tilde\varphi^{abc}+\omega_5\, \tilde\alpha^d_a\epsilon^{abc}\tilde\varphi_{dbc}\right)\ .
		\end{split}
	\ee
The second commutator \eqref{appqt,pt} yields the same result, with $\alpha\leftrightarrow\tilde\alpha$. The two choices are alternative and equivalent, in fact by choosing the first \eqref{appq,p} we have $q\sim\alpha$, $p\sim\tilde\alpha$, while the second \eqref{appqt,pt} corresponds to $\tilde q\sim\tilde\alpha$, $\tilde p\sim\alpha$. Something similar also happens in Maxwell theory \cite{Bertolini:2020hgr}.



\end{document}